\documentclass[10pt,aps,pra,twocolumn,superscriptaddress,floatfix]{revtex4-2}

\usepackage{amsmath,amssymb,graphicx,bm}
\usepackage{hyperref}
\hypersetup{hidelinks}
\makeatletter
\def\@bibdataout@aps{%
 \immediate\write\@bibdataout{%
  @CONTROL{%
   apsrev42Control%
   \longbibliography@sw{%
    ,author="48",editor="1",pages="0",title="0",year="1"%
   }{%
    ,author="48",editor="1",pages="0",title="",year="1"%
   }%
  }%
 }%
 \if@filesw
  \immediate\write\@auxout{\string\citation{apsrev42Control}}%
 \fi
}%
\makeatother

\begin{document}

\title{Single-impurity polarons in hard-core lattice bosons at low and intermediate fillings}

\author{Chao Zhang}
\email{chaozhang@ahnu.edu.cn}
\affiliation{Department of Physics, Anhui Normal University, Wuhu, 241002, China}


\begin{abstract}
We investigate a single mobile impurity in a two-dimensional hard-core
Bose--Hubbard bath at low and intermediate fillings and determine how polaronic
dressing evolves with bath filling for impurity--bath couplings ranging from
weak to strong and ultimately to the two-component hard-core limit.
Using large-scale, sign-problem-free worm-algorithm quantum Monte Carlo simulations, we
extract momentum-space quasiparticle properties from the impurity Green's
function and resolve the accompanying real-space bath rearrangement from an
imaginary-time-averaged impurity-centered correlator.
We also vary the impurity hopping $t_{\rm imp}$ to assess how reduced mobility
modifies dressing in the strong-coupling regime.
For the fillings accessible at each coupling, the impurity remains a dressed quasiparticle whose
ground-state energy, effective mass, and residue vary smoothly with filling
$n_{\rm b}$.
In real space, increasing $n_{\rm b}$ strengthens the short-range depletion
while shifting the dominant response toward the impurity.
In the two-component hard-core limit $U_{\rm ib}/t_{\rm b}\!\to\!\infty$, the
large-distance recovery of the cumulative density deformation exhibits only
weak filling dependence over the range considered here, whereas short-range core
indicators continue to evolve.
Our results quantitatively characterize strongly dressed polarons in a
correlated, compressible lattice bath and resolve how filling and impurity
mobility modify the near-core response and spatial extent of the dressing
cloud.
\end{abstract}

\maketitle

\section{Introduction}

The concept of a polaron---a mobile impurity dressed by excitations of its
environment---plays a central role in our understanding of quasiparticles in
interacting many-body systems~\cite{Landau1933,
Frohlich1950,
Feynman1955,
Holstein1959,
PhysRevB.104.035143,ckbn-jp9t,PhysRevB.109.165119}.
Since its original formulation in the context of electrons coupled to lattice
vibrations, polaron physics has provided a unifying framework for how
microscopic interactions renormalize the effective properties of a particle,
including its energy, mass, and coherence.
In recent years, this paradigm has gained renewed interest in ultracold atomic
gases, where impurity problems can be realized 
with high tunability and probed with unprecedented precision through
radio-frequency spectroscopy and related techniques~\cite{JorgensenPRL2016,
HuPRL2016,ScazzaZaccanti2022}.

A particularly rich setting for polaron physics arises when an impurity is
immersed in a bosonic environment.
Both continuum Bose gases and bosonic lattice baths have been explored extensively~\cite{ArdilaGiorgini2015,Ardila2020,DuttaMueller2013,DingSciPost2023},
revealing dressing mechanisms ranging from weak-coupling polarons dressed by
long-wavelength density/phase fluctuations to strongly correlated bound states.
Recent work has further addressed one- and two-dimensional settings,
strong-coupling and Efimov-related regimes, lattice square geometries,
hard-core bosonic baths, related platforms such as Rydberg and ionic
impurities, and structured-impurity extensions~\cite{GrusdtNJP2017,
SchmidtEnss2022,ChristianenPRA2022,DingSciPost2023,
CamargoPRL2018,AstrakharchikCommPhys2021,PenaArdilaCamacho2025}.
In weakly interacting or dilute bosonic baths, theoretical descriptions based on
Bogoliubov theory and related Fr\"ohlich or field-theoretic impurity--BEC descriptions
often provide accurate and intuitive
pictures~\cite{TemperePRB2009,RathSchmidtPRA2013,ShashiPRA2014,
GrusdtSciRep2015}.
However, when the bath itself is strongly correlated, these approaches are no
longer controlled, and the impurity problem must be addressed from a fully
many-body perspective.

The Bose--Hubbard model offers a paradigmatic lattice platform in which strong
correlation effects emerge naturally~\cite{JakschPRL1998, PhysRevB.40.546, PhysRevB.75.134302, CapogrossoSansonePRA2008}.
In particular, the hard-core limit enforces a strict on-site occupancy constraint
$n_i\in\{0,1\}$ for bath bosons, dramatically altering their local and collective
properties.
While hard-core bosons remain compressible and superfluid below unit filling, the
constraint suppresses local density fluctuations and induces nontrivial
correlations.
At finite densities of both components, vacancy motion in this
hard-core model can mediate strong nondissipative drag between component currents
through polaronic correlations~\cite{KielyZhangMuellerPRA2025}.
An impurity moving in such a background therefore might experience a different environment from softcore bosonic baths~\cite{chaoletter,chaoPRB}.

Despite its fundamental interest, a mobile impurity embedded in a
two-dimensional hard-core bosonic lattice at incommensurate filling
$n_{\mathrm b}<1$ has received relatively limited attention. This regime offers
a clean setting in which the on-site constraint $n_i\in\{0,1\}$ maximizes
local exclusion effects while the bath remains gapless and compressible away
from commensurate filling. It therefore helps disentangle the consequences of
strong local constraints from those of genuine incompressibility. A basic question is then how the impurity dressing evolves as the repulsive
impurity--bath coupling $U_{\mathrm{ib}}$ increases: does the system remain
continuously connected to the weak-coupling polaron, or can the depletion
cloud undergo a more pronounced restructuring? 

Recent variational work investigated the spectral and quasiparticle
properties of an impurity in a two-dimensional hard-core-boson condensate and
predicted a strong suppression of the repulsive-polaron residue close to unit
filling at strong coupling~\cite{Santiago-García_2024}.  Here we complement
that work with a sign-problem-free QMC study at the low and intermediate
fillings investigated for each coupling, combining the energy, effective mass, and
residue with impurity-centered real-space diagnostics and reduced-mobility data;
we do not address the asymptotic near-unit-filling regime.

In this work, we address these questions by studying a single mobile impurity in
the two-dimensional hard-core Bose--Hubbard model below unit filling.
Using large-scale, sign-problem-free worm-algorithm quantum Monte Carlo (QMC)
simulations~\cite{ProkofevJETP1998,secondworm, PhysRevA.81.053622,Lingua_2018}, we determine both momentum-space quasiparticle properties---the
ground-state energy, effective mass, and quasiparticle residue---and real-space
bath rearrangements encoded in the impurity-centered correlator
$C_{\rm ib}(\mathbf {r})$ and the cumulative bath-density deformation
$\Delta N(R)$.
In addition, we vary the impurity hopping $t_{\rm imp}$ to assess how reduced
bare mobility modifies the strong-coupling dressing cloud.
By combining these complementary diagnostics, we provide a unified
characterization of impurity dressing across a wide range of bath fillings and
impurity--bath coupling strengths, including the two-component hard-core limit.

Within the coupling-dependent filling range explored here,
increasing $n_{\rm b}$ smoothly enhances the mass renormalization and suppresses
$Z_0$ for all impurity--bath couplings studied.  In the strong finite-coupling
and two-component hard-core datasets, the dominant real-space response also
shifts toward the impurity.  The impurity remains a coherent dressed
quasiparticle over the filling ranges explored here. Together, the
momentum-space and impurity-centered real-space QMC diagnostics quantitatively
resolve how bath filling and bare impurity mobility modify the near-core
response and spatial extent of the dressing cloud.

The remainder of the paper is organized as follows.
Section~\ref{model} introduces the two-component Bose--Hubbard Hamiltonian. In Sec.~\ref{sec:methods} we briefly summarize the worm-algorithm QMC approach and define the observables used throughout.
Specifically, Sec.~\ref{subsec:GF_quasiparticle} describes how the impurity Green's function is measured and how the quasiparticle properties (ground-state energy, effective mass, and residue) are extracted, while Sec.~\ref{subsec:obs_cib} presents the impurity-centered correlator $C_{\rm ib}(\mathbf r)$ and the derived real-space diagnostics.
Our main results for polaron properties in a hard-core bath are discussed in Sec.~\ref{sec:results_basic}, and the two-component hard-core limit is analyzed in detail in Sec.~\ref{subsec:hard-core}.
We additionally present a comparison between different impurity hoppings $t_{\rm imp}$ to quantify the impact of reduced impurity mobility on the strong-coupling dressing in Sec.~\ref{subsec:timp05}.
Finally, we conclude in Sec.~\ref{sec:conclusion}.

\begin{figure}[tbp]
    \centering
    \includegraphics[width=\columnwidth]{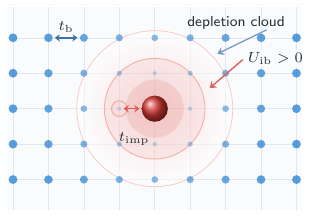}
    \caption{\textbf{Schematic of a repulsive impurity in a low-filling hard-core bath.}
    A single mobile impurity (red sphere) moves by nearest-neighbor hopping
    $t_{\rm imp}$ on a square lattice and interacts repulsively with a hard-core Bose--Hubbard bath
    ($U_{\rm ib}>0$).
    Blue dots schematically encode the local bath density at $n_{\rm b}<1$;
    bath particles hop between neighboring lattice sites with amplitude $t_{\rm b}$.
    The impurity expels nearby bath particles, producing a depletion cloud around
    its trajectory.
    }
    \label{fig:schematic_model}
\end{figure}

\section{Model}
\label{model}

We study a single mobile impurity immersed in a two-dimensional
hard-core Bose--Hubbard bath on a square lattice.
The system is described by the Hamiltonian
\begin{align}
H = &
-t_{\mathrm{imp}} \!\!\sum_{\langle i,j\rangle}
   (a_{i}^\dagger a_j + \mathrm{H.c.})
+ U_{\mathrm{ib}}\sum_i n_{\mathrm{imp},i} n_{\mathrm{b},i} \nonumber \\
& -t_{\mathrm{b}} \!\!\sum_{\langle i,j\rangle}
   (b_i^\dagger b_j + \mathrm{H.c.})
- \mu_{\mathrm{b}}\sum_i n_{\mathrm{b},i},
\label{eq:Hamiltonian}
\end{align}
where $b_i^\dagger$ ($a_i^\dagger$) creates a bath (impurity) boson
on lattice site $i$,
and $n_{\mathrm{b},i}=b_i^\dagger b_i$,
$n_{\mathrm{imp},i}=a_i^\dagger a_i$ are the corresponding
on-site number operators.

The bath bosons obey the hard-core constraint
\begin{equation}
n_{\mathrm b,i}\in\{0,1\},
\end{equation}
which forbids double occupancy and induces strong local correlations.
The bath hopping amplitude $t_{\mathrm{b}}$ sets the energy scale
($t_{\mathrm{b}}=1$ throughout).
The impurity hopping is taken as
$t_{\mathrm{imp}}=1$ or $0.5$.
The local impurity--bath coupling $U_{\mathrm{ib}}$ is the key tuning
parameter and $U_{\mathrm{ib}}>0$ corresponds to a repulsive impurity--bath coupling.

The bath chemical potential $\mu_{\mathrm{b}}$ controls the bath filling
\begin{equation}
n_{\mathrm{b}} =\frac{1}{L^2} \sum_i \langle n_{\mathrm{b},i} \rangle ,
\end{equation}
with $L$ the system size and throughout this work we focus on the compressible regime
$n_{\mathrm{b}}<1$.
A single impurity is enforced by the constraint
$\sum_i n_{\mathrm{imp},i}=1$,
i.e., the impurity sector is simulated at fixed particle number one.

Figure~\ref{fig:schematic_model} provides a schematic view of the physical situation studied in this work.
A single impurity occupies a lattice site and interacts locally and repulsively with the compressible hard-core bath ($U_{\rm{ib}}>0$), thereby pushing bath particles away from its vicinity.
At low filling this response remains smooth and spatially extended, and it may be viewed as a depletion cloud dressing the impurity motion.
In the following sections we quantify this dressing by extracting the impurity dispersion and quasiparticle properties ($E_{p}(0)$, $m^*/m_0$, and $Z_{0}$) and by characterizing the bath density response around the impurity.

\vspace{6pt}
\noindent

\section{Method and observables}
\label{sec:methods}

We employ a sign-problem-free two-species worm-algorithm quantum Monte Carlo
(QMC) simulation in the path-integral representation, which samples the
worldline configurations of bath and impurity bosons in an enlarged
configuration space. All simulations are performed on square lattices
with periodic boundary conditions and inverse temperatures $\beta=L=40$.


Our analysis combines momentum-space quasiparticle diagnostics and real-space
impurity-centered correlators.
Momentum-space properties are obtained from the impurity Green's function
$G_{\mathrm{imp}}(\mathbf k,\tau)$, from which we extract the polaron energy
$E_{\mathrm p}(\mathbf k)$, quasiparticle residue $Z_{\mathbf k}$, and the
effective mass $m^*$ from the small-$\mathbf k$ dispersion.
Real-space structure is quantified by an imaginary-time-averaged impurity-centered bath-density profile
$G_{\rm ic}(\mathbf r)$, the associated impurity-centered correlator
$C_{\rm ib}(\mathbf r)$, its radially averaged correlator
$C_{\rm ib}(R)$, and scalar measures derived from $C_{\rm ib}(R)$ that
characterize the strength and extent of the dressing cloud.
Statistical uncertainties are estimated using a jackknife analysis over
independent Monte Carlo blocks.

\subsection{Impurity Green's function and quasiparticle properties}
\label{subsec:GF_quasiparticle}

We extract momentum-space quasiparticle properties of the impurity from its
single-particle Green's function,
\begin{equation}
G_{\mathrm{imp}}(\mathbf r,\tau)
=
\Bigl\langle
a_{\mathrm{imp}}(\mathbf r,\tau)\,
a_{\mathrm{imp}}^{\dagger}(\mathbf 0,0)
\Bigr\rangle,
\label{eq:Gimp_rt}
\end{equation}
where $\mathbf r$ denotes the lattice displacement (equivalently, the site index
$i$ used in Sec.~\ref{model} under periodic boundary conditions), and
$\tau\in[0,\beta]$ is the imaginary-time separation.
Within the multi-species worm algorithm, $G_{\mathrm{imp}}(\mathbf r,\tau)$ is
measured from the statistics of the impurity worm-end separation in the
off-diagonal sector~\cite{ProkofevJETP1998,secondworm, PhysRevA.81.053622,Lingua_2018}.

The momentum-resolved Green's function is obtained by a discrete Fourier
transform,
\begin{equation}
G_{\mathrm{imp}}(\mathbf k,\tau)
=
\sum_{\mathbf r}
e^{-i\mathbf k\cdot \mathbf r}\,
G_{\mathrm{imp}}(\mathbf r,\tau),
\label{eq:Gimp_kt}
\end{equation}
with lattice momenta $\mathbf k=(2\pi n_x/L,\,2\pi n_y/L)$ with $n_x$ and $n_y$ integers.

At sufficiently low temperature, the asymptotic behavior is dominated
by the lowest polaron state at momentum $\mathbf k$,
\begin{equation}
G_{\mathrm{imp}}(\mathbf k,\tau)
\simeq
Z_{\mathbf k}\,
e^{-E_{\mathrm p}(\mathbf k)\,\tau}.
\label{eq:Gimp_asympt}
\end{equation}
This defines the polaron dispersion $E_{\mathrm p}(\mathbf k)$ and the
quasiparticle residue $Z_{\mathbf k}$.

\paragraph*{Ground-state energy and residue.}
At $\mathbf k=\mathbf 0$, we fit $\ln G_{\mathrm{imp}}(\mathbf 0,\tau)$ to a
linear form within a $\tau$ interval where a single-exponential decay is clearly
observed. The slope yields the polaron ground-state energy $E_{\mathrm p}(\mathbf 0)$,
while the intercept determines the residue $Z_{\mathbf 0}$ via
Eq.~\eqref{eq:Gimp_asympt}. 

\paragraph*{Effective mass.}
To determine the effective mass we first extract the polaron dispersion
$E_{\mathrm p}(\mathbf k)$ from the long-$\tau$ decay of
$G_{\mathrm{imp}}(\mathbf k,\tau)$ [Eq.~\eqref{eq:Gimp_asympt}] at the lowest
accessible lattice momenta. 
We report the mass renormalization relative to the bare (small-$k$) lattice mass
$m_0\equiv (2t_{\mathrm{imp}})^{-1}$,
\begin{equation}
\frac{m^*}{m_0}
=
\frac{2t_{\mathrm{imp}}}{\left.\dfrac{\partial^2 E_{\mathrm p}(\mathbf k)}
{\partial k_\alpha^2}\right|_{\mathbf k=\mathbf 0}},
\qquad (\alpha=x\ \text{or}\ y),
\label{eq:mstar}
\end{equation}
where the second derivative is evaluated from the quadratic fit to the
lowest-momentum data. 

\begin{figure*}[t]
\centering
\includegraphics[width=\textwidth]{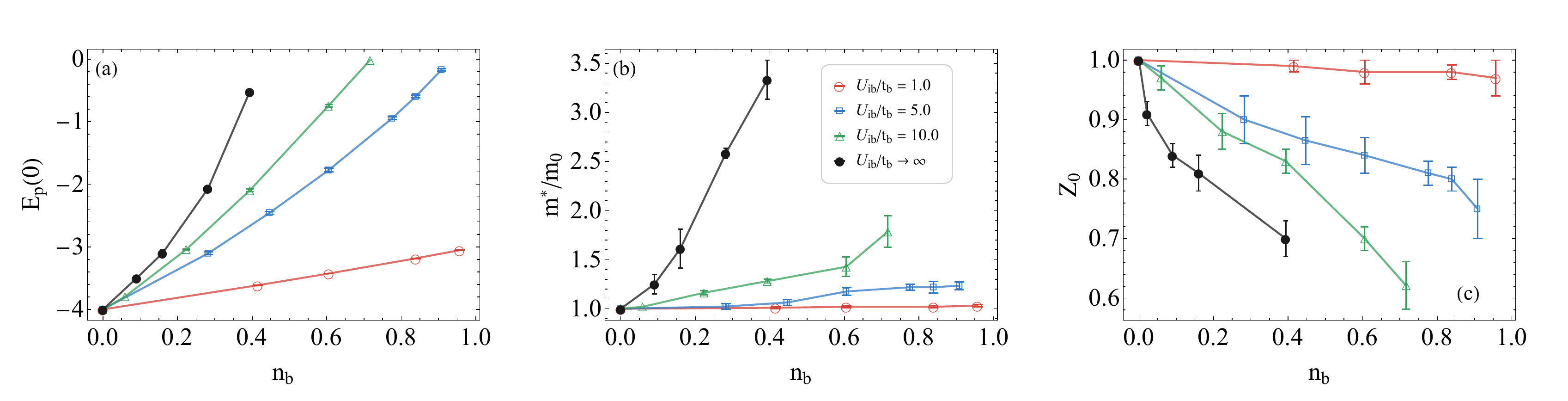}
\caption{\textbf{Quasiparticle properties of a single impurity in a hard-core bosonic bath at incommensurate filling for $t_{\rm imp}=t_{\rm b}=1.0$.}
Shown as functions of the bath filling $n_{\mathrm b}$ for several values of the impurity--bath coupling strength $U_{\rm ib}/t_{\rm b}$.
(a)~Ground-state energy $E_p(0)$ of the impurity.
(b)~Effective mass ratio $m^*/m_0$, extracted from the curvature of the impurity dispersion at small momentum.
(c)~Quasiparticle residue $Z_0$ at zero momentum.
}
\label{fig:fig1}
\end{figure*}

\subsection{Impurity-centered correlator and derived observables}
\label{subsec:obs_cib}

A central goal of this work is to quantify how strongly and over what spatial
extent the bath is distorted by a single mobile impurity.
We therefore define an \emph{impurity-conditioned} bath-density profile in
real space and construct all real-space diagnostics from it.

\paragraph*{Impurity-centered bath-density profile and correlator.}
In the worldline representation the impurity visits different sites along imaginary time.
For the single impurity, this defines the imaginary-time-averaged spatial weight
\begin{equation}
\begin{aligned}
w(\mathbf r_{\rm imp})
&=
\frac{1}{\beta}\int_{0}^{\beta} d\tau\;
n_{\rm imp}(\mathbf r_{\rm imp},\tau),\\
\sum_{\mathbf r_{\rm imp}} w(\mathbf r_{\rm imp})&=1 .
\end{aligned}
\label{eq:w_imp}
\end{equation}
Here $\bar n_{\rm b}(\mathbf r)\equiv \beta^{-1}\int_{0}^{\beta} d\tau\,
n_{\rm b}(\mathbf r,\tau)$ denotes the imaginary-time-averaged bath occupancy.
We define the impurity-centered (conditioned) bath-density profile
\begin{equation}
G_{\rm ic}(\mathbf r)
=
\Bigg\langle
\sum_{\mathbf r_{\rm imp}} w(\mathbf r_{\rm imp})\,
\bar n_{\rm b}(\mathbf r_{\rm imp}+\mathbf r)
\Bigg\rangle,
\label{eq:Gic_r}
\end{equation}
where $\mathbf r$ is a displacement measured with the minimum-image convention
under periodic boundary conditions.
The corresponding uniform bath background is the bath filling $n_{\rm b}$
defined in Sec.~\ref{model}.
Here $w(\mathbf r_{\rm imp})$ and $\bar n_{\rm b}(\mathbf r)$ are understood as
time-averaged quantities, while $\langle\cdots\rangle$ denotes the ensemble average.
The \emph{imaginary-time-averaged impurity-centered correlator} is
then defined as
\begin{equation}
C_{\rm ib}(\mathbf r)=G_{\rm ic}(\mathbf r)-n_{\rm b}.
\label{eq:Cib_r}
\end{equation}
Thus, $C_{\rm ib}(\mathbf r)$ is built from separately time-averaged
impurity and bath occupancies.
For repulsive $U_{\rm ib}>0$ one typically finds $C_{\rm ib}(\mathbf r)<0$ near the
impurity (depletion). $G_{\rm ic}(\mathbf r)$ is the conditioned bath-density profile itself, and
$C_{\rm ib}(\mathbf r)$ is the corresponding background-subtracted
correlator.

\paragraph*{Radially averaged impurity-centered correlator.}
To suppress lattice anisotropy and obtain a smooth profile, we compute the
radially averaged impurity-centered correlator
\begin{equation}
C_{\rm ib}(R)
=
\frac{1}{N_R}\sum_{|\mathbf r|=R} C_{\rm ib}(\mathbf r),
\label{eq:Cib_R}
\end{equation}
Here $R=\sqrt{r_x^2+r_y^2}$ is the Euclidean distance under
the minimum-image convention, and $N_R$ counts the lattice vectors satisfying
$|\mathbf r|=R$.  Thus $C_{\rm ib}(R)$ averages only over equal-distance sites,
without combining distinct distances into integer radial bins.

\paragraph*{Cumulative bath-density deformation.}
From $C_{\rm ib}(R)$ we construct the cumulative bath-density deformation within
radius $R$,
\begin{equation}
\Delta N(R)
=
\sum_{R'\le R} N_{R'}\, C_{\rm ib}(R').
\label{eq:DeltaN_R}
\end{equation}
$\Delta N(R)$ therefore gives the net change of bath particle number
within radius $R$ centered at the impurity.
By construction, $\Delta N(R\!\to\!\infty)=0$ (up to statistical
noise), because $C_{\rm ib}(\mathbf r)$ is defined relative to the global bath
filling $n_{\rm b}$ within the same ensemble.  Equivalently, the complete-lattice
sum obeys $\sum_{\mathbf r}C_{\rm ib}(\mathbf r)=0$.  The behavior at finite $R$
reveals the local deformation created by the impurity, whereas the return to
zero follows from this background-subtraction sum rule and cannot serve as
independent evidence for polaron stability.

\paragraph*{Shell-averaged dressing strength.}
To characterize the overall magnitude of the impurity-induced distortion
independent of its sign, we define the shell-averaged dressing strength
\begin{equation}
\mathcal D_{\rm pol}
\equiv
\sum_{R} \big|C_{\rm ib}(R)\big|.
\label{eq:Dpol}
\end{equation}
Here $C_{\rm ib}(R)$ is the radially averaged impurity-centered
correlator at shell radius $R$.
Because the shell multiplicity $N_R$ has already been divided out in $C_{\rm ib}(R)$,
$\mathcal D_{\rm pol}$ measures the shell-averaged amplitude of the impurity-induced
response rather than the total displaced particle number.
Accordingly, $\mathcal D_{\rm pol}$ is an operational shell-averaged measure built from
the radial profile.
The fully integrated particle-number information is instead encoded in $\Delta N(R)$.
This definition keeps each distance shell on equal footing and is therefore well
suited for tracking how the radial profile reorganizes across a smooth crossover.

\paragraph*{Shell-based polaron radius.}
A complementary operational measure of the spatial extent is defined from
the second moment of $|C_{\rm ib}(R)|$,
\begin{equation}
\xi_{\rm pol}^2
\equiv
\frac{\sum_{R} R^2\,\big|C_{\rm ib}(R)\big|}{\sum_{R} \big|C_{\rm ib}(R)\big|}.
\label{eq:xipol}
\end{equation}
Here $\xi_{\rm pol}$ is the shell-based polaron radius:
it measures where the weight of the shell-averaged response $|C_{\rm ib}(R)|$ is concentrated in space.
Here $R$ retains the exact Euclidean radius, so $\xi_{\rm pol}$
is a second moment over exact lattice-distance shells.
A smaller $\xi_{\rm pol}$ means that the dominant response has moved closer to the impurity,
even if the cumulative bath-density deformation encoded in $\Delta N(R)$ continues to evolve.

\paragraph*{On-site contrast.}
Finally, the most local indicator of dressing is the on-site contrast
\begin{equation}
C_{\rm ib}(0) \equiv C_{\rm ib}(R=0),
\label{eq:Cib0}
\end{equation}
which probes the zero-displacement value of the impurity-centered correlator.

In the notation used throughout, $C_{\rm ib}(\mathbf r)$ is the
impurity-centered correlator, $C_{\rm ib}(R)$ its radially
averaged correlator,
$\Delta N(R)$ the cumulative bath-density deformation, $\mathcal D_{\rm pol}$ the
shell-averaged dressing strength, $\xi_{\rm pol}$ the shell-based polaron radius,
and $C_{\rm ib}(0)$ the on-site contrast.

\section{Polaron properties in a hard-core bath}
\label{sec:results_basic}

\subsection{Single-impurity quasiparticle properties}
\label{subsec:qp}

We begin by characterizing the impurity through its momentum-space
quasiparticle properties.
Figure~\ref{fig:fig1} summarizes the evolution of the ground-state energy $E_p(0)$,
effective mass $m^*/m_0$, and quasiparticle residue $Z_0$ as functions of the bath filling
$n_{\mathrm b}$ for different impurity--bath coupling strengths.
For $t_{\rm imp}=t_{\rm b}$, the quasiparticle datasets extend up to
$n_{\rm b}\simeq0.96$, $0.90$, $0.72$, and $0.39$ for
$U_{\rm ib}/t_{\rm b}=1$, $5$, $10$, and $\infty$, respectively.

As shown in Fig.~\ref{fig:fig1}(a), the impurity ground-state energy
$E_p(0)$ increases monotonically with bath filling for all
$U_{\rm ib}/t_{\rm b}$.
This trend reflects the increasing cost of creating a local distortion in a
denser bath subject to the hard-core constraint.
Importantly, $E_p(0)$ evolves smoothly with $n_{\mathrm b}$ over the entire
filling range studied.

Figure~\ref{fig:fig1}(b) shows the effective mass ratio $m^*/m_0$ extracted
from the low-momentum dispersion of the impurity Green’s function.
With increasing $n_{\mathrm b}$, the effective mass grows, signaling reduced
impurity mobility due to enhanced dressing by bath density fluctuations.
This effect is particularly pronounced in the hard-core limit, where the
absence of double occupancy strongly constrains bath rearrangements around
the impurity.
Nevertheless, even in this extreme case the mass enhancement remains finite
at the highest fillings accessible in the hard-core dataset, consistent with a heavy
but mobile polaron.

The quasiparticle residue $Z_0$, shown in
Fig.~\ref{fig:fig1}(c), provides a complementary measure of impurity dressing.
As either $n_{\mathrm b}$ or $U_{\rm ib}/t_{\rm b}$ is increased, $Z_0$ decreases
continuously.
Notably, $Z_0$ remains nonzero for all fillings and impurity-bath coupling strengths
studied here, including the hard-core-limit case, demonstrating that
within the explored range the impurity retains coherent quasiparticle
character.

Taken together, Figs.~\ref{fig:fig1}(a)--(c) show a smooth evolution
from weaker to stronger dressing as the bath density increases.  Over the
filling range investigated at each coupling, the extracted mass remains
finite and the residue remains nonzero.  These data support a coherent polaron
over the explored range.

\begin{figure*}[t]
\centering
\includegraphics[width=0.78\textwidth, trim={2cm 0.0cm 15.0cm 0.1cm}, clip]{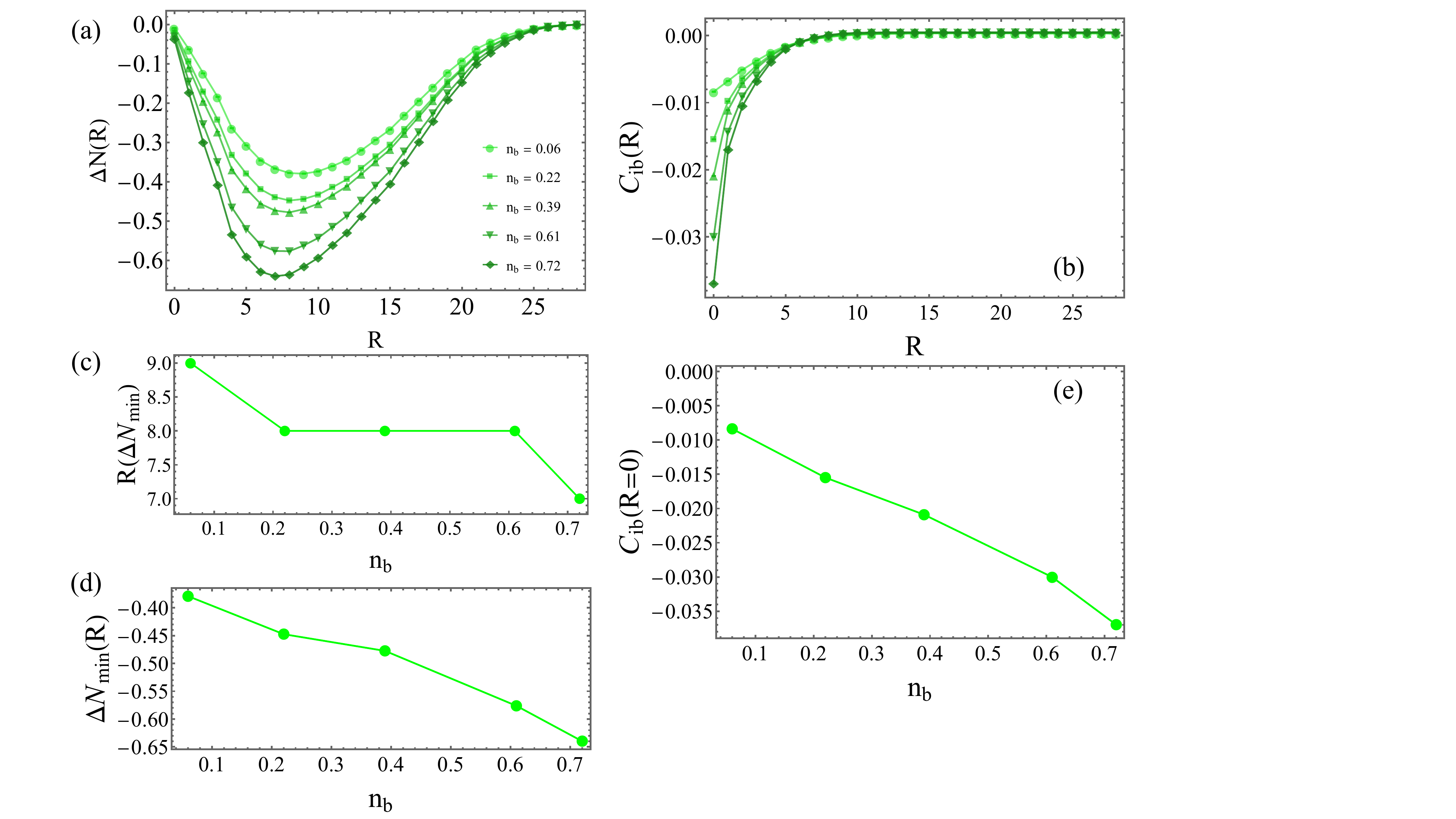}
\caption{\textbf{Real-space dressing cloud at strong but finite repulsion $U_{\rm ib}/t_{\rm b}=10.0$ for $t_{\rm imp}=t_{\rm b}=1.0$.}
Data are shown for bath fillings $0.06\le n_{\mathrm b}\le 0.72$.
All panels are constructed from the impurity-centered correlator $C_{\rm ib}(\mathbf r)$ and its radially averaged correlator
$C_{\rm ib}(R)$.
(a) Cumulative bath-density deformation $\Delta N(R)$.
(b) Radially averaged impurity-centered correlator $C_{\rm ib}(R)$.
(c,d) The location $R(\Delta N_{\min})$ of the minimum of $\Delta N(R)$ and the
corresponding minimum value $\Delta N_{\min}$.
(e) On-site contrast $C_{\rm ib}(0)$ as a function of bath filling $n_{\mathrm b}$.}
\label{fig:Uib10_realspace}
\end{figure*}

\subsection{Real-space structure at strong finite repulsion}
\label{subsec:realspace_Uib10}

To complement the momentum-space quasiparticle diagnostics, we now resolve the
microscopic structure of impurity dressing in real space.
Our central object is the impurity-centered correlator
$C_{\rm ib}(\mathbf r)$; we then perform radial averaging to obtain $C_{\rm ib}(R)$, from which all
derived measures below are constructed.
Figure~\ref{fig:Uib10_realspace} summarizes the results at strong repulsion
$U_{\rm ib}/t_{\rm b}=10.0$ as a function of bath filling $n_{\mathrm b}$.

The cumulative bath-density deformation
$\Delta N(R)$ in Fig.~\ref{fig:Uib10_realspace}(a)
exhibits a pronounced negative minimum for all fillings, demonstrating the
formation of a depletion cloud around the impurity.
With increasing $n_{\mathrm b}$ the minimum depth $\Delta N_{\min}$ becomes
progressively more negative, i.e., the impurity expels more bath weight from its
vicinity.
At large distances, $\Delta N(R)$ returns smoothly to zero by construction, as
required by the background-subtraction sum rule discussed in
Sec.~\ref{subsec:obs_cib}.

The radially averaged impurity-centered correlator $C_{\rm ib}(R)$ in Fig.~\ref{fig:Uib10_realspace}(b)
shows a small negative response at small $R$ and decays to zero within
$R\sim 7$--$8$.
The magnitude of the near-origin depletion increases monotonically with
$n_{\mathrm b}$, reflecting stronger local exclusion in a denser hard-core bath.
The decay indicates that the dressing cloud is spatially compact at
$U_{\rm ib}/t_{\rm b}=10.0$, while its overall strength can still vary substantially.

To summarize the deformation profile in Fig.~\ref{fig:Uib10_realspace}(a), we
track two robust features extracted from $\Delta N(R)$: the radius of its minimum
$R(\Delta N_{\min})$ and the minimum value $\Delta N_{\min}$ at this $R$
[Figs.~\ref{fig:Uib10_realspace}(c,d)].
A notable trend is that $R(\Delta N_{\min})$ stays stable across
$0.06\le n_{\mathrm b}\le 0.72$.
Thus, increasing $n_{\mathrm b}$ does not substantially shift the characteristic
radius at which $\Delta N(R)$ is most negative.
Instead, the dominant effect of filling is on the \emph{amplitude} of the local
response: $\Delta N_{\min}$ becomes more negative as $n_{\mathrm b}$ increases.
This is consistent with the behavior of the on-site contrast
$C_{\rm ib}(0)$ [Fig.~\ref{fig:Uib10_realspace}(e)], whose magnitude also grows with
filling.
Although $C_{\rm ib}(0)$ remains numerically small, its monotonic trend shows that the
extra dressing acquired at larger $n_{\mathrm b}$ is concentrated in the near-core region
rather than in a broad expansion of the cloud.

Within the investigated interval $0.06\le n_{\mathrm b}\le0.72$,
$\Delta N_{\min}$ and $C_{\rm ib}(0)$ evolve smoothly and monotonically, without
a plateau or abrupt change.  The real-space response therefore shows a gradual
increase of the short-range dressing amplitude. This trend accompanies the
smooth mass and residue data in Sec.~\ref{subsec:qp}.

\subsection{Real-space structure in the hard-core limit}
\label{subsec:hard-core}

\begin{figure*}[t]
\centering
\includegraphics[width=0.78\textwidth, trim={2cm 0.0cm 13.0cm 0.1cm}, clip]{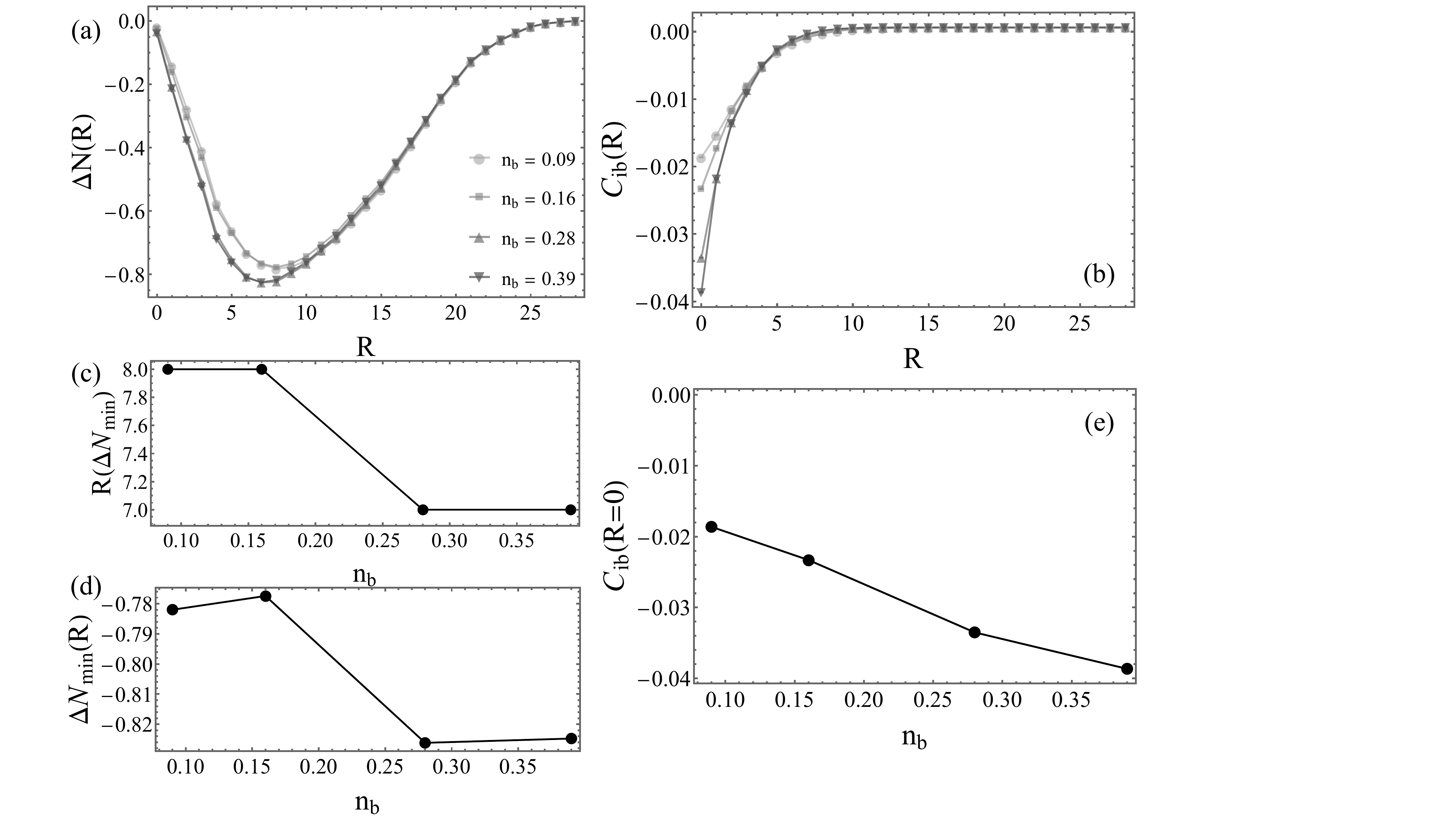}
\caption{\textbf{Real-space dressing cloud in the two-component hard-core limit for $t_{\rm imp}=t_{\rm b}=1.0$.}
Data are shown for $U_{\rm ib}/t_{\rm b}\!\to\!\infty$ (both species
hard-core) at bath fillings $n_{\mathrm b}=0.09,\,0.16,\,0.28,$ and $0.39$.
All panels are constructed from the impurity-centered correlator
$C_{\rm ib}(\mathbf r)$ and its radially averaged correlator $C_{\rm ib}(R)$.
(a) Cumulative bath-density deformation $\Delta N(R)$.
(b) Radially averaged impurity-centered correlator $C_{\rm ib}(R)$.
(c,d) Location of the minimum $R(\Delta N_{\min})$ and the minimum value $\Delta N_{\min}$.
(e) On-site contrast $C_{\rm ib}(0)$ as a function of bath filling $n_{\mathrm b}$.
}
\label{fig:hard-core_realspace}
\end{figure*}

We finally consider the two-component hard-core limit, where both the impurity and
the bath satisfy the on-site constraint $n_i\in\{0,1\}$ and the impurity--bath
repulsion is effectively infinite, $U_{\rm ib}/t_{\rm b}\to\infty$.

Figure~\ref{fig:hard-core_realspace}(a) shows the cumulative bath-density deformation
$\Delta N(R)$ for several bath fillings.
For all fillings considered, $\Delta N(R)$ develops a clear negative minimum at a
finite radius and then recovers smoothly toward zero at large $R$, as required
by the same background-subtraction sum rule.

Over $0.09\le n_{\mathrm b}\le0.39$, the large-$R$ portions of the
$\Delta N(R)$ curves overlap closely and therefore exhibit weak filling
dependence at the available resolution.  By contrast, increasing $n_{\mathrm b}$
changes the depth of $\Delta N_{\min}$ and the short-range magnitude of
$C_{\rm ib}(R)$.  This comparison indicates that the observed filling
dependence is concentrated mainly in the short-distance response.

To quantify the dressing profile we plot the position of the minimum,
$R(\Delta N_{\min})$, and its depth $\Delta N_{\min}$ in
Fig.~\ref{fig:hard-core_realspace}(c,d).
We find that $R(\Delta N_{\min})$ depends only weakly on $n_{\mathrm b}$,
whereas $\Delta N_{\min}$ becomes more negative with increasing $n_{\mathrm b}$.

Figure~\ref{fig:hard-core_realspace}(e) plots $C_{\rm ib}(0)$ as a
function of $n_{\mathrm b}$.  As defined in Sec.~\ref{subsec:obs_cib}, this is an
imaginary-time-averaged impurity-centered correlator, not an equal-time
double-occupancy correlator.  It can therefore be small and nonzero even in the
two-component hard-core limit.  Its magnitude increases from roughly $0.02$ to
$0.04$ across the plotted fillings, consistent with the stronger near-core
response in the other panels.

\begin{figure}[tbp]
\centering
\includegraphics[width=\linewidth]{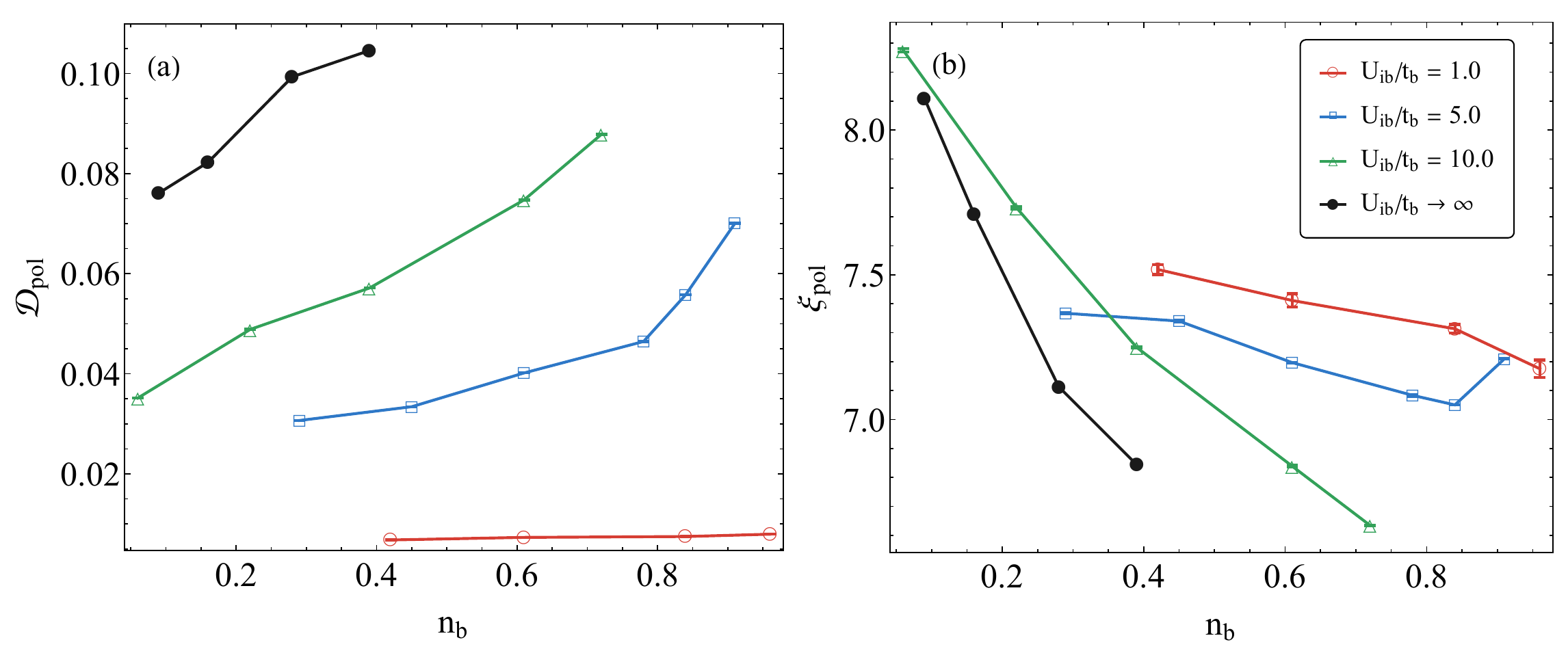}
\caption{\textbf{Strength and extent of the polaron dressing cloud from the radially averaged impurity-centered correlator for $t_{\rm imp}=t_{\rm b}=1.0$.}
Real-space diagnostics of the impurity-induced bath deformation in the
compressible regime $n_{\mathrm b}<1$, extracted from the radially averaged correlator
$C_{\rm ib}(R)$ (see Sec.~\ref{sec:methods}).
(a) Shell-averaged dressing strength $\mathcal D_{\mathrm{pol}}$ as a function of bath filling $n_{\rm b}$.
(b) Shell-based polaron radius $\xi_{\mathrm{pol}}$ as a function of bath filling $n_{\rm b}$.
}
\label{fig:dpol_xipol}
\end{figure}

\begin{figure*}[t]
\centering
\includegraphics[width=\linewidth]{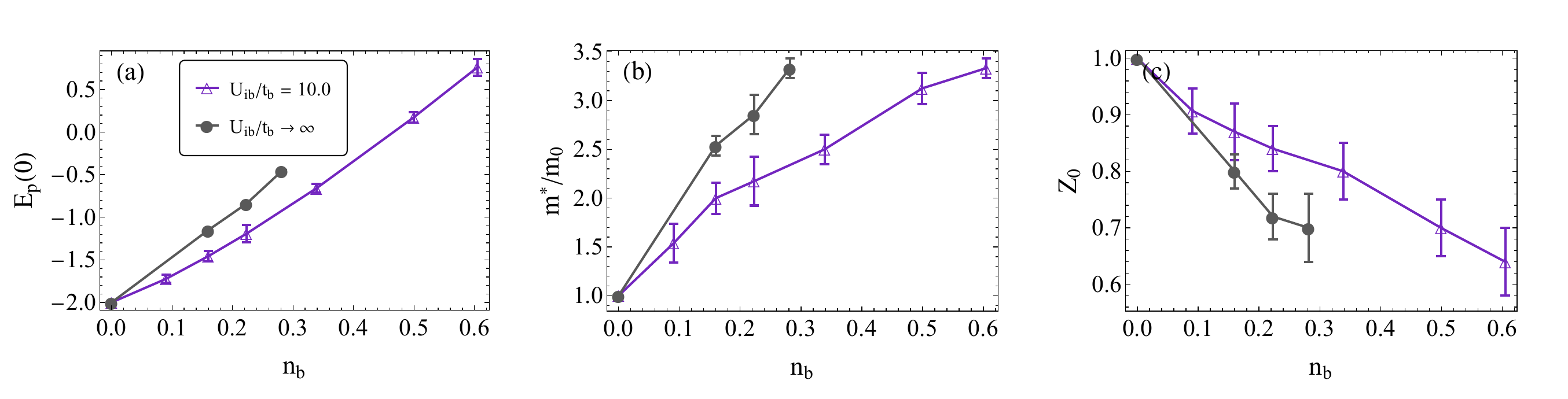}
\caption{\textbf{Impurity quasiparticle properties for reduced impurity hopping
$t_{\mathrm{imp}}=0.5$ at strong impurity--bath coupling.}
Ground-state energy $E_{\mathrm p}(0)$(a), effective mass $m^*/m_0$(b), and quasiparticle
residue $Z_0$(c) as functions of bath filling $n_{\mathrm b}$ for two interaction
strengths: strong but finite coupling $U_{\rm ib}/t_{\rm b}=10.0$ (purple) and the
two-component hard-core limit $U_{\rm ib}/t_{\rm b}\!\to\!\infty$ (gray).
}
\label{fig:timp05_quasiparticle}
\end{figure*}

\subsection{Shell-averaged dressing strength and shell-based polaron radius}
\label{subsec:dpol_xipol}

Momentum-space quantities such as the effective mass and quasiparticle residue
directly characterize the impurity mobility and coherence, but they do not
explicitly show how the bath is reorganized in real space.
A complementary geometric summary is provided by the impurity-centered correlator.
Since our primary real-space observable is the radially averaged profile
$C_{\rm ib}(R)$, it is useful to compress this profile into two scalar measures
that characterize its typical amplitude and spatial extent.
The definitions of the shell-averaged dressing strength $\mathcal D_{\rm pol}$
and the shell-based polaron radius $\xi_{\rm pol}$ are given in
Sec.~\ref{subsec:obs_cib}; here we focus on their physical interpretation and
their evolution with bath filling.

Within this framework, $\mathcal D_{\rm pol}$ measures the overall magnitude of
the radial response across distance shells: larger values indicate that the
impurity induces a stronger local rearrangement of the bath, independent of its sign.
The length scale $\xi_{\rm pol}$ measures where this shell-averaged weight is concentrated.
A decrease of $\xi_{\rm pol}$ therefore indicates that the dominant response is
pulled closer to the impurity rather than distributed over a broader range of $R$.

Figure~\ref{fig:dpol_xipol}(a) shows that $\mathcal D_{\rm pol}$ increases
monotonically with $n_{\mathrm b}$ for all couplings considered.
Physically, as the bath becomes denser, more bath weight is available near the
impurity and the hard-core constraint makes local exclusion more effective, so the
short-range depletion sharpens and the shell-averaged amplitude grows.

In contrast, Fig.~\ref{fig:dpol_xipol}(b) shows that the shell-based polaron radius
$\xi_{\rm pol}$ decreases with increasing $n_{\mathrm b}$.
Below unit filling the bath remains compressible, increasing density primarily enhances
short-range exclusion and local rearrangement, which naturally compresses the effective
dressing length scale.

The opposite trends of $\mathcal D_{\rm pol}$ and $\xi_{\rm pol}$ reveal a clear
separation between the magnitude and the spatial extent of polaronic dressing:
increasing $n_{\mathrm b}$ amplifies the shell-averaged deformation amplitude while shifting
the dominant response toward shorter distances.
Across the investigated two-component hard-core fillings,
$\mathcal D_{\rm pol}$ increases with $n_{\mathrm b}$ while $\xi_{\rm pol}$ decreases,
indicating a stronger but more spatially concentrated response around the
impurity.

\subsection{Effect of reduced impurity mobility}
\label{subsec:timp05}

So far we have focused on an impurity with hopping $t_{\mathrm{imp}}=1$.
To isolate the role of impurity mobility, we now reduce the hopping to
$t_{\mathrm{imp}}=0.5$ while keeping the bath hard-core and at incommensurate
fillings $n_{\rm b}<1$.
We consider this reduced-mobility impurity in two representative impurity--bath
coupling regimes: a strong but finite repulsion $U_{\rm ib}/t_{\rm b}=10.0$ and
the two-component hard-core limit $U_{\rm ib}/t_{\rm b}\to\infty$.
Decreasing $t_{\mathrm{imp}}$ doubles the bare lattice mass
$m_0=(2t_{\mathrm{imp}})^{-1}$ and lowers the impurity kinetic-energy scale,
thereby amplifying polaronic dressing at fixed interaction strength.
Importantly, for $n_{\rm b}<1$ the bath remains compressible and its Hamiltonian
is unchanged; thus varying $t_{\mathrm{imp}}$ primarily tunes the impurity
mobility.
This provides a controlled knob to disentangle kinetic effects from
impurity--bath coupling and to quantify the competition between impurity motion
and local impurity--bath exclusion.

Figure~\ref{fig:timp05_quasiparticle} summarizes the impurity quasiparticle
properties for reduced impurity hopping $t_{\mathrm{imp}}=0.5$ at strong
impurity--bath repulsion, comparing finite coupling
$U_{\rm ib}/t_{\rm b}=10.0$ with the two-component hard-core limit
$U_{\rm ib}/t_{\rm b}\rightarrow\infty$.
In addition, the $t_{\mathrm{imp}}=0.5$ results can be contrasted with our
baseline $t_{\mathrm{imp}}=1.0$ data (Sec.~\ref{subsec:qp}) for the mobility- and
coherence-sensitive observables: reducing $t_{\mathrm{imp}}$ doubles the bare
lattice mass $m_0=(2t_{\mathrm{imp}})^{-1}$ and lowers the impurity kinetic scale,
thereby enhancing bath-induced dressing at fixed bath parameters and fixed
$U_{\rm ib}/t_{\rm b}$.

The polaron ground-state energy $E_{\mathrm p}(0)$
[Fig.~\ref{fig:timp05_quasiparticle}(a)] evolves smoothly with bath filling for
both coupling limits.
For $t_{\mathrm{imp}}=0.5$, $E_{\mathrm p}(0)$ increases monotonically with
$n_{\mathrm b}$, reflecting the growing energetic cost of propagating an impurity
through a denser background.

The effective-mass renormalization $m^*/m_0$
[Fig.~\ref{fig:timp05_quasiparticle}(b)] exhibits a pronounced enhancement with
increasing filling, demonstrating progressive suppression of impurity mobility
by bath-induced dressing.
At fixed $t_{\mathrm{imp}}$, the mass enhancement is stronger in
the two-component hard-core limit than at $U_{\rm ib}/t_{\rm b}=10.0$, reflecting
the most restrictive local constraint.
Crucially, relative to the $t_{\mathrm{imp}}=1.0$ case the mass renormalization
is amplified across the entire filling range studied: lowering $t_{\mathrm{imp}}$
provides an efficient route to enhance polaronic dressing without modifying the
bath itself.
Nevertheless, in both interaction limits $m^*/m_0$ varies continuously with
$n_{\mathrm b}$.

Consistently, the quasiparticle residue $Z_0$
[Fig.~\ref{fig:timp05_quasiparticle}(c)] decreases as $n_{\mathrm b}$ increases,
signaling a gradual loss of coherent impurity motion.
The suppression is more pronounced in the hard-core limit, again reflecting
stronger short-range impurity--bath couplings.
The extracted $Z_0$ remains finite throughout the investigated parameter
range, supporting a coherent dressed quasiparticle within that range.

Overall, at fixed strong repulsion the two knobs---strengthening the local
constraint (finite $U_{\rm ib}$ versus hard-core) and reducing the impurity
mobility ($t_{\mathrm{imp}}=1.0$ versus $0.5$)---act in the same qualitative direction:
they enhance the mass renormalization and suppress the
residue, i.e., they make the polaron heavier and less coherent.
At the same time, all quasiparticle observables remain smooth functions of
$n_{\mathrm b}$ over the respective investigated ranges, indicating a continuous
strengthening of polaronic dressing over the parameter range resolved here.

\begin{figure}[!t]
\centering
\includegraphics[width=\linewidth]{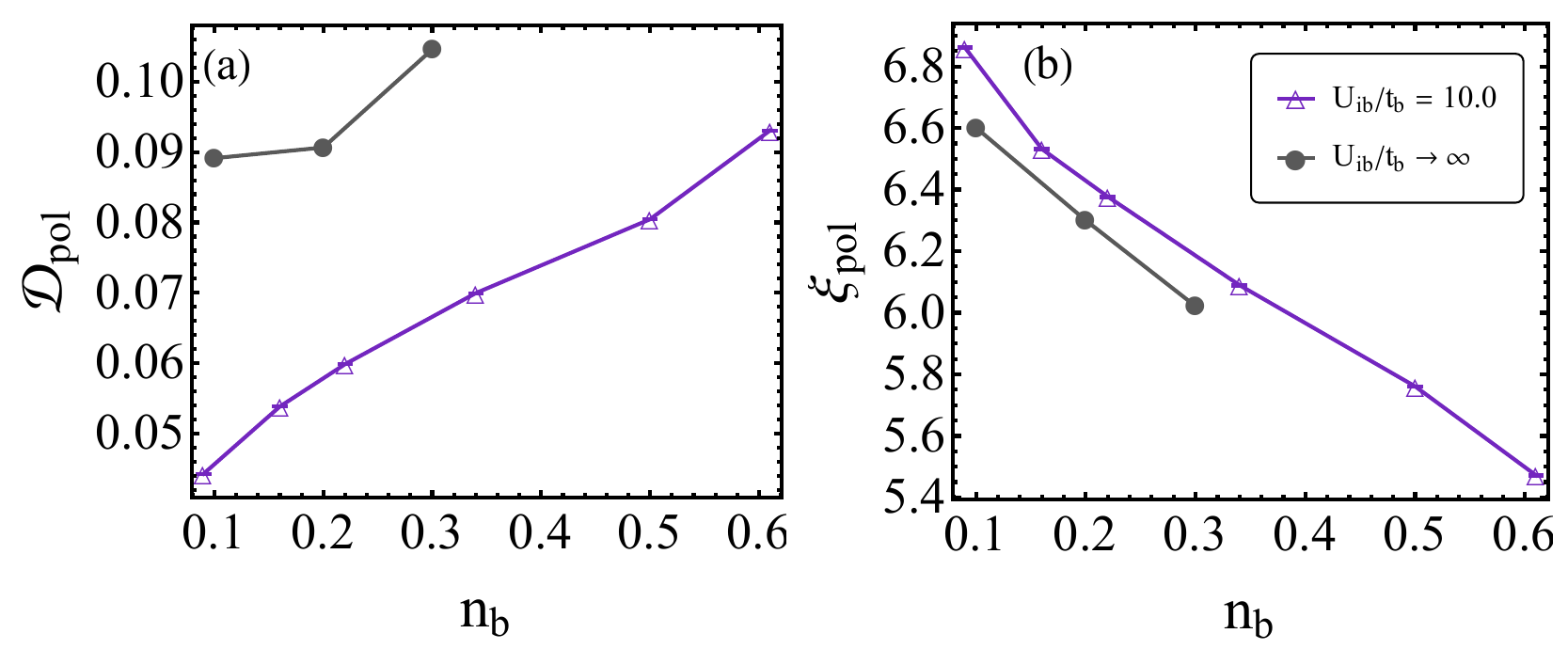}
\caption{\textbf{Shell-averaged dressing strength and shell-based polaron radius} for reduced impurity mobility $t_{\mathrm{imp}}=0.5$.
Real-space scalars extracted from $C_{\rm ib}(R)$ for strong but finite coupling
$U_{\rm ib}/t_{\rm b}=10.0$ (purple) and the two-component hard-core limit
$U_{\rm ib}/t_{\rm b}\!\to\!\infty$ (gray).
The purple series extends to $n_{\mathrm b}=0.60$, whereas the hard-core reference is
available up to $n_{\mathrm b}=0.30$.
(a) Shell-averaged dressing strength $\mathcal D_{\mathrm{pol}}$ as a function of bath filling $n_{\rm b}$.
(b) Shell-based polaron radius $\xi_{\mathrm{pol}}$ as a function of bath filling $n_{\rm b}$.
}
\label{fig:timp05_cloud}
\end{figure}

The real-space consequences of reduced impurity mobility are summarized in
Fig.~\ref{fig:timp05_cloud} in terms of the
shell-averaged dressing strength $\mathcal D_{\mathrm{pol}}$ and the
shell-based polaron radius $\xi_{\mathrm{pol}}$ for reduced impurity mobility $t_{\mathrm{imp}}=0.5$.
For $t_{\mathrm{imp}}=0.5$, $\mathcal D_{\mathrm{pol}}$ increases with bath filling
while $\xi_{\mathrm{pol}}$ decreases, for both the strong finite-coupling and hard-core
results shown. The added dressing therefore accumulates mainly in the near-core region,
again indicating a response that becomes stronger without spreading further out.

Comparing with the baseline $t_{\mathrm{imp}}=1.0$ results clarifies the role of
reduced mobility. Lowering $t_{\mathrm{imp}}$ makes the impurity slower, so the induced
deformation is less able to propagate over many shells.
Taken together, these results show that lowering $t_{\mathrm{imp}}$ primarily
reshuffles the balance between core strength and spatial extent: the response
becomes more localized in real space while remaining smooth as a function of bath
filling.

\subsection{Experimental realization}

The setting considered here can be approached with a dilute second
atomic component immersed in a two-dimensional bosonic gas in a square optical
lattice.  A large bath interaction-to-hopping ratio suppresses doublons and
approximates the hard-core regime, while a local spin flip or dilute doping can
prepare a single impurity.  State-dependent lattices and interspecies Feshbach
resonances provide control of $t_{\rm imp}/t_{\rm b}$ and
$U_{\rm ib}/t_{\rm b}$~\cite{FukuharaNatPhys2013,JorgensenPRL2016,
HuPRL2016}.  Radio-frequency spectroscopy can probe the polaron energy and the
coherent spectral weight associated with $Z_0$, while wave-packet expansion or
momentum-resolved probes can constrain the effective mass.  Species-resolved
quantum-gas microscopy can measure the equal-time bath-density profile
conditioned on the simultaneously observed impurity position.  Centering and
averaging repeated snapshots on that position provides an experimental
analogue of the real-space dressing cloud, although this observable is not
identical to the imaginary-time-averaged $C_{\rm ib}(\mathbf r)$ defined here.
The hard-core and $U_{\rm ib}/t_{\rm b}\!\to\!\infty$ cases correspond
experimentally to large but finite interaction-to-hopping ratios with strongly
suppressed double occupancy.

\section{Conclusion}
\label{sec:conclusion}

We studied a single mobile impurity in a two-dimensional hard-core
Bose--Hubbard bath using sign-problem-free QMC, combining the ground-state
energy, effective mass, and quasiparticle residue with impurity-centered
real-space diagnostics.  Our conclusions are restricted to the
coupling-dependent low- and intermediate-filling ranges explored here.  Across
these ranges, increasing filling produces a smooth increase of the energy and
mass renormalization and a smooth reduction of the residue, while the extracted
residue remains nonzero. 

The real-space measurements show that increasing $n_{\mathrm b}$ strengthens
the short-range depletion response and shifts the dominant shell-averaged
weight toward the impurity.  Across the investigated fillings in the
two-component hard-core limit, the large-distance part of $\Delta N(R)$ shows
only weak filling dependence, whereas the near-core response exhibits a clearer
filling dependence.  By construction, the full-lattice sum
of $C_{\rm ib}(\mathbf r)$ vanishes because the global bath filling of the same
ensemble is subtracted.  The eventual return of $\Delta N(R)$ to zero therefore
follows from a sum rule and is not used as independent evidence for polaron
stability.

Reducing the impurity hopping to $t_{\mathrm{imp}}=0.5$ enhances mass
renormalization and suppresses the residue over the investigated range, while the
real-space response becomes more concentrated near the core.  Taken together,
these sign-problem-free QMC results quantitatively characterize strongly
dressed polarons in a compressible hard-core lattice bath and show how filling
and impurity mobility modify the strength and spatial extent of the dressing
cloud.


\begin{acknowledgments}
This work was supported by the National Natural Science Foundation
of China under Grant No.~12204173.
\end{acknowledgments}

\bibliography{impurity}

\begin{thebibliography}{37}%
\makeatletter
\providecommand \@ifxundefined [1]{%
 \@ifx{#1\undefined}
}%
\providecommand \@ifnum [1]{%
 \ifnum #1\expandafter \@firstoftwo
 \else \expandafter \@secondoftwo
 \fi
}%
\providecommand \@ifx [1]{%
 \ifx #1\expandafter \@firstoftwo
 \else \expandafter \@secondoftwo
 \fi
}%
\providecommand \natexlab [1]{#1}%
\providecommand \enquote  [1]{``#1''}%
\providecommand \bibnamefont  [1]{#1}%
\providecommand \bibfnamefont [1]{#1}%
\providecommand \citenamefont [1]{#1}%
\providecommand \href@noop [0]{\@secondoftwo}%
\providecommand \href [0]{\begingroup \@sanitize@url \@href}%
\providecommand \@href[1]{\@@startlink{#1}\@@href}%
\providecommand \@@href[1]{\endgroup#1\@@endlink}%
\providecommand \@sanitize@url [0]{\catcode `\\12\catcode `\$12\catcode
  `\&12\catcode `\#12\catcode `\^12\catcode `\_12\catcode `\%12\relax}%
\providecommand \@@startlink[1]{}%
\providecommand \@@endlink[0]{}%
\providecommand \url  [0]{\begingroup\@sanitize@url \@url }%
\providecommand \@url [1]{\endgroup\@href {#1}{\urlprefix }}%
\providecommand \urlprefix  [0]{URL }%
\providecommand \Eprint [0]{\href }%
\providecommand \doibase [0]{https://doi.org/}%
\providecommand \selectlanguage [0]{\@gobble}%
\providecommand \bibinfo  [0]{\@secondoftwo}%
\providecommand \bibfield  [0]{\@secondoftwo}%
\providecommand \translation [1]{[#1]}%
\providecommand \BibitemOpen [0]{}%
\providecommand \bibitemStop [0]{}%
\providecommand \bibitemNoStop [0]{.\EOS\space}%
\providecommand \EOS [0]{\spacefactor3000\relax}%
\providecommand \BibitemShut  [1]{\csname bibitem#1\endcsname}%
\let\auto@bib@innerbib\@empty
\bibitem [{\citenamefont {Landau}(1933)}]{Landau1933}%
  \BibitemOpen
  \bibfield  {author} {\bibinfo {author} {\bibfnamefont {L.~D.}\ \bibnamefont
  {Landau}},\ }\bibfield  {title} {\bibinfo {title} {On the motion of electrons
  in crystal lattices},\ }\href
  {https://ui.adsabs.harvard.edu/abs/1933PhyZS...3..664L} {\bibfield  {journal}
  {\bibinfo  {journal} {Physikalische Zeitschrift der Sowjetunion}\ }\textbf
  {\bibinfo {volume} {3}},\ \bibinfo {pages} {664} (\bibinfo {year}
  {1933})}\BibitemShut {NoStop}%
\bibitem [{\citenamefont {Fr\"ohlich}\ \emph {et~al.}(1950)\citenamefont
  {Fr\"ohlich}, \citenamefont {Pelzer},\ and\ \citenamefont
  {Zienau}}]{Frohlich1950}%
  \BibitemOpen
  \bibfield  {author} {\bibinfo {author} {\bibfnamefont {H.}~\bibnamefont
  {Fr\"ohlich}}, \bibinfo {author} {\bibfnamefont {H.}~\bibnamefont {Pelzer}},\
  and\ \bibinfo {author} {\bibfnamefont {S.}~\bibnamefont {Zienau}},\
  }\bibfield  {title} {\bibinfo {title} {Properties of slow electrons in polar
  materials},\ }\href {https://doi.org/10.1080/14786445008521782} {\bibfield
  {journal} {\bibinfo  {journal} {Philosophical Magazine}\ }\textbf {\bibinfo
  {volume} {41}},\ \bibinfo {pages} {221} (\bibinfo {year} {1950})}\BibitemShut
  {NoStop}%
\bibitem [{\citenamefont {Feynman}(1955)}]{Feynman1955}%
  \BibitemOpen
  \bibfield  {author} {\bibinfo {author} {\bibfnamefont {R.~P.}\ \bibnamefont
  {Feynman}},\ }\bibfield  {title} {\bibinfo {title} {Slow electrons in a polar
  crystal},\ }\href {https://doi.org/10.1103/PhysRev.97.660} {\bibfield
  {journal} {\bibinfo  {journal} {Physical Review}\ }\textbf {\bibinfo {volume}
  {97}},\ \bibinfo {pages} {660} (\bibinfo {year} {1955})}\BibitemShut
  {NoStop}%
\bibitem [{\citenamefont {Holstein}(1959)}]{Holstein1959}%
  \BibitemOpen
  \bibfield  {author} {\bibinfo {author} {\bibfnamefont {T.}~\bibnamefont
  {Holstein}},\ }\bibfield  {title} {\bibinfo {title} {Studies of polaron
  motion: Part i. the molecular-crystal model},\ }\href
  {https://doi.org/10.1016/0003-4916(59)90002-8} {\bibfield  {journal}
  {\bibinfo  {journal} {Annals of Physics}\ }\textbf {\bibinfo {volume} {8}},\
  \bibinfo {pages} {325} (\bibinfo {year} {1959})}\BibitemShut {NoStop}%
\bibitem [{\citenamefont {Zhang}\ \emph {et~al.}(2021)\citenamefont {Zhang},
  \citenamefont {Prokof'ev},\ and\ \citenamefont
  {Svistunov}}]{PhysRevB.104.035143}%
  \BibitemOpen
  \bibfield  {author} {\bibinfo {author} {\bibfnamefont {C.}~\bibnamefont
  {Zhang}}, \bibinfo {author} {\bibfnamefont {N.~V.}\ \bibnamefont
  {Prokof'ev}},\ and\ \bibinfo {author} {\bibfnamefont {B.~V.}\ \bibnamefont
  {Svistunov}},\ }\bibfield  {title} {\bibinfo {title}
  {Peierls/su-schrieffer-heeger polarons in two dimensions},\ }\href
  {https://doi.org/10.1103/PhysRevB.104.035143} {\bibfield  {journal} {\bibinfo
   {journal} {Phys. Rev. B}\ }\textbf {\bibinfo {volume} {104}},\ \bibinfo
  {pages} {035143} (\bibinfo {year} {2021})}\BibitemShut {NoStop}%
\bibitem [{\citenamefont {Zhang}(2025)}]{ckbn-jp9t}%
  \BibitemOpen
  \bibfield  {author} {\bibinfo {author} {\bibfnamefont {C.}~\bibnamefont
  {Zhang}},\ }\bibfield  {title} {\bibinfo {title} {Comprehensive study of bond
  bipolaron superconductivity on the triangular lattice},\ }\href
  {https://doi.org/10.1103/PhysRevB.112.174520} {\bibfield  {journal} {\bibinfo
   {journal} {Phys. Rev. B}\ }\textbf {\bibinfo {volume} {112}},\ \bibinfo
  {pages} {174520} (\bibinfo {year} {2025})}\BibitemShut {NoStop}%
\bibitem [{\citenamefont {Zhang}(2024)}]{PhysRevB.109.165119}%
  \BibitemOpen
  \bibfield  {author} {\bibinfo {author} {\bibfnamefont {C.}~\bibnamefont
  {Zhang}},\ }\bibfield  {title} {\bibinfo {title} {Light polarons with
  electron-phonon coupling},\ }\href
  {https://doi.org/10.1103/PhysRevB.109.165119} {\bibfield  {journal} {\bibinfo
   {journal} {Phys. Rev. B}\ }\textbf {\bibinfo {volume} {109}},\ \bibinfo
  {pages} {165119} (\bibinfo {year} {2024})}\BibitemShut {NoStop}%
\bibitem [{\citenamefont {J\o{}rgensen}\ \emph {et~al.}(2016)\citenamefont
  {J\o{}rgensen}, \citenamefont {Wacker}, \citenamefont {Skalmstang},
  \citenamefont {Parish}, \citenamefont {Levinsen}, \citenamefont
  {Christensen}, \citenamefont {Bruun},\ and\ \citenamefont
  {Arlt}}]{JorgensenPRL2016}%
  \BibitemOpen
  \bibfield  {author} {\bibinfo {author} {\bibfnamefont {N.~B.}\ \bibnamefont
  {J\o{}rgensen}}, \bibinfo {author} {\bibfnamefont {L.}~\bibnamefont
  {Wacker}}, \bibinfo {author} {\bibfnamefont {K.~T.}\ \bibnamefont
  {Skalmstang}}, \bibinfo {author} {\bibfnamefont {M.~M.}\ \bibnamefont
  {Parish}}, \bibinfo {author} {\bibfnamefont {J.}~\bibnamefont {Levinsen}},
  \bibinfo {author} {\bibfnamefont {R.~S.}\ \bibnamefont {Christensen}},
  \bibinfo {author} {\bibfnamefont {G.~M.}\ \bibnamefont {Bruun}},\ and\
  \bibinfo {author} {\bibfnamefont {J.~J.}\ \bibnamefont {Arlt}},\ }\bibfield
  {title} {\bibinfo {title} {Observation of attractive and repulsive polarons
  in a bose-einstein condensate},\ }\href
  {https://doi.org/10.1103/PhysRevLett.117.055302} {\bibfield  {journal}
  {\bibinfo  {journal} {Phys. Rev. Lett.}\ }\textbf {\bibinfo {volume} {117}},\
  \bibinfo {pages} {055302} (\bibinfo {year} {2016})}\BibitemShut {NoStop}%
\bibitem [{\citenamefont {Hu}\ \emph {et~al.}(2016)\citenamefont {Hu},
  \citenamefont {Van~de Graaff}, \citenamefont {Kedar}, \citenamefont {Corson},
  \citenamefont {Cornell},\ and\ \citenamefont {Jin}}]{HuPRL2016}%
  \BibitemOpen
  \bibfield  {author} {\bibinfo {author} {\bibfnamefont {M.-G.}\ \bibnamefont
  {Hu}}, \bibinfo {author} {\bibfnamefont {M.~J.}\ \bibnamefont {Van~de
  Graaff}}, \bibinfo {author} {\bibfnamefont {D.}~\bibnamefont {Kedar}},
  \bibinfo {author} {\bibfnamefont {J.~P.}\ \bibnamefont {Corson}}, \bibinfo
  {author} {\bibfnamefont {E.~A.}\ \bibnamefont {Cornell}},\ and\ \bibinfo
  {author} {\bibfnamefont {D.~S.}\ \bibnamefont {Jin}},\ }\bibfield  {title}
  {\bibinfo {title} {Bose polarons in the strongly interacting regime},\ }\href
  {https://doi.org/10.1103/PhysRevLett.117.055301} {\bibfield  {journal}
  {\bibinfo  {journal} {Phys. Rev. Lett.}\ }\textbf {\bibinfo {volume} {117}},\
  \bibinfo {pages} {055301} (\bibinfo {year} {2016})}\BibitemShut {NoStop}%
\bibitem [{\citenamefont {Scazza}\ \emph {et~al.}(2022)\citenamefont {Scazza},
  \citenamefont {Zaccanti}, \citenamefont {Massignan}, \citenamefont {Parish},\
  and\ \citenamefont {Levinsen}}]{ScazzaZaccanti2022}%
  \BibitemOpen
  \bibfield  {author} {\bibinfo {author} {\bibfnamefont {F.}~\bibnamefont
  {Scazza}}, \bibinfo {author} {\bibfnamefont {M.}~\bibnamefont {Zaccanti}},
  \bibinfo {author} {\bibfnamefont {P.}~\bibnamefont {Massignan}}, \bibinfo
  {author} {\bibfnamefont {M.~M.}\ \bibnamefont {Parish}},\ and\ \bibinfo
  {author} {\bibfnamefont {J.}~\bibnamefont {Levinsen}},\ }\bibfield  {title}
  {\bibinfo {title} {Repulsive fermi and bose polarons in quantum gases},\
  }\bibfield  {journal} {\bibinfo  {journal} {Atoms}\ }\textbf {\bibinfo
  {volume} {10}},\ \href {https://doi.org/10.3390/atoms10020055}
  {10.3390/atoms10020055} (\bibinfo {year} {2022})\BibitemShut {NoStop}%
\bibitem [{\citenamefont {Pe{\~n}a~Ardila}\ and\ \citenamefont
  {Giorgini}(2015)}]{ArdilaGiorgini2015}%
  \BibitemOpen
  \bibfield  {author} {\bibinfo {author} {\bibfnamefont {L.~A.}\ \bibnamefont
  {Pe{\~n}a~Ardila}}\ and\ \bibinfo {author} {\bibfnamefont {S.}~\bibnamefont
  {Giorgini}},\ }\bibfield  {title} {\bibinfo {title} {Impurity in a
  {B}ose--{E}instein condensate: Study of the attractive and repulsive branch
  using quantum {M}onte {C}arlo methods},\ }\href
  {https://doi.org/10.1103/PhysRevA.92.033612} {\bibfield  {journal} {\bibinfo
  {journal} {Physical Review A}\ }\textbf {\bibinfo {volume} {92}},\ \bibinfo
  {pages} {033612} (\bibinfo {year} {2015})}\BibitemShut {NoStop}%
\bibitem [{\citenamefont {Pe{\~n}a~Ardila}\ \emph {et~al.}(2020)\citenamefont
  {Pe{\~n}a~Ardila}, \citenamefont {Astrakharchik},\ and\ \citenamefont
  {Giorgini}}]{Ardila2020}%
  \BibitemOpen
  \bibfield  {author} {\bibinfo {author} {\bibfnamefont {L.~A.}\ \bibnamefont
  {Pe{\~n}a~Ardila}}, \bibinfo {author} {\bibfnamefont {G.~M.}\ \bibnamefont
  {Astrakharchik}},\ and\ \bibinfo {author} {\bibfnamefont {S.}~\bibnamefont
  {Giorgini}},\ }\bibfield  {title} {\bibinfo {title} {Strong-coupling {B}ose
  polarons in a two-dimensional gas},\ }\href
  {https://doi.org/10.1103/PhysRevResearch.2.023405} {\bibfield  {journal}
  {\bibinfo  {journal} {Physical Review Research}\ }\textbf {\bibinfo {volume}
  {2}},\ \bibinfo {pages} {023405} (\bibinfo {year} {2020})}\BibitemShut
  {NoStop}%
\bibitem [{\citenamefont {Dutta}\ and\ \citenamefont
  {Mueller}(2013)}]{DuttaMueller2013}%
  \BibitemOpen
  \bibfield  {author} {\bibinfo {author} {\bibfnamefont {S.}~\bibnamefont
  {Dutta}}\ and\ \bibinfo {author} {\bibfnamefont {E.~J.}\ \bibnamefont
  {Mueller}},\ }\bibfield  {title} {\bibinfo {title} {Variational study of
  polarons and bipolarons in a one-dimensional bose lattice gas in both the
  superfluid and the mott-insulator regimes},\ }\href
  {https://doi.org/10.1103/PhysRevA.88.053601} {\bibfield  {journal} {\bibinfo
  {journal} {Phys. Rev. A}\ }\textbf {\bibinfo {volume} {88}},\ \bibinfo
  {pages} {053601} (\bibinfo {year} {2013})}\BibitemShut {NoStop}%
\bibitem [{\citenamefont {Ding}\ \emph {et~al.}(2023)\citenamefont {Ding},
  \citenamefont {Dom{\'\i}nguez-Castro}, \citenamefont {Julku}, \citenamefont
  {Camacho-Guardian},\ and\ \citenamefont {Bruun}}]{DingSciPost2023}%
  \BibitemOpen
  \bibfield  {author} {\bibinfo {author} {\bibfnamefont {S.}~\bibnamefont
  {Ding}}, \bibinfo {author} {\bibfnamefont {G.~A.}\ \bibnamefont
  {Dom{\'\i}nguez-Castro}}, \bibinfo {author} {\bibfnamefont {A.}~\bibnamefont
  {Julku}}, \bibinfo {author} {\bibfnamefont {A.}~\bibnamefont
  {Camacho-Guardian}},\ and\ \bibinfo {author} {\bibfnamefont {G.~M.}\
  \bibnamefont {Bruun}},\ }\bibfield  {title} {\bibinfo {title} {Polarons and
  bipolarons in a two-dimensional square lattice},\ }\href
  {https://doi.org/10.21468/SciPostPhys.14.6.143} {\bibfield  {journal}
  {\bibinfo  {journal} {SciPost Phys.}\ }\textbf {\bibinfo {volume} {14}},\
  \bibinfo {pages} {143} (\bibinfo {year} {2023})}\BibitemShut {NoStop}%
\bibitem [{\citenamefont {Grusdt}\ \emph {et~al.}(2017)\citenamefont {Grusdt},
  \citenamefont {Astrakharchik},\ and\ \citenamefont {Demler}}]{GrusdtNJP2017}%
  \BibitemOpen
  \bibfield  {author} {\bibinfo {author} {\bibfnamefont {F.}~\bibnamefont
  {Grusdt}}, \bibinfo {author} {\bibfnamefont {G.~E.}\ \bibnamefont
  {Astrakharchik}},\ and\ \bibinfo {author} {\bibfnamefont {E.}~\bibnamefont
  {Demler}},\ }\bibfield  {title} {\bibinfo {title} {Bose polarons in ultracold
  atoms in one dimension: beyond the {Fr{\"o}}hlich paradigm},\ }\href
  {https://doi.org/10.1088/1367-2630/aa8a2e} {\bibfield  {journal} {\bibinfo
  {journal} {New Journal of Physics}\ }\textbf {\bibinfo {volume} {19}},\
  \bibinfo {pages} {103035} (\bibinfo {year} {2017})}\BibitemShut {NoStop}%
\bibitem [{\citenamefont {Schmidt}\ and\ \citenamefont
  {Enss}(2022)}]{SchmidtEnss2022}%
  \BibitemOpen
  \bibfield  {author} {\bibinfo {author} {\bibfnamefont {R.}~\bibnamefont
  {Schmidt}}\ and\ \bibinfo {author} {\bibfnamefont {T.}~\bibnamefont {Enss}},\
  }\bibfield  {title} {\bibinfo {title} {Self-stabilized bose polarons},\
  }\href {https://doi.org/10.21468/SciPostPhys.13.3.054} {\bibfield  {journal}
  {\bibinfo  {journal} {SciPost Phys.}\ }\textbf {\bibinfo {volume} {13}},\
  \bibinfo {pages} {054} (\bibinfo {year} {2022})}\BibitemShut {NoStop}%
\bibitem [{\citenamefont {Christianen}\ \emph {et~al.}(2022)\citenamefont
  {Christianen}, \citenamefont {Cirac},\ and\ \citenamefont
  {Schmidt}}]{ChristianenPRA2022}%
  \BibitemOpen
  \bibfield  {author} {\bibinfo {author} {\bibfnamefont {A.}~\bibnamefont
  {Christianen}}, \bibinfo {author} {\bibfnamefont {J.~I.}\ \bibnamefont
  {Cirac}},\ and\ \bibinfo {author} {\bibfnamefont {R.}~\bibnamefont
  {Schmidt}},\ }\bibfield  {title} {\bibinfo {title} {Bose polaron and the
  {E}fimov effect: A gaussian-state approach},\ }\href
  {https://doi.org/10.1103/PhysRevA.105.053302} {\bibfield  {journal} {\bibinfo
   {journal} {Phys. Rev. A}\ }\textbf {\bibinfo {volume} {105}},\ \bibinfo
  {pages} {053302} (\bibinfo {year} {2022})}\BibitemShut {NoStop}%
\bibitem [{\citenamefont {Camargo}\ \emph {et~al.}(2018)\citenamefont
  {Camargo}, \citenamefont {Schmidt}, \citenamefont {Whalen}, \citenamefont
  {Ding}, \citenamefont {Woehl}, \citenamefont {Yoshida}, \citenamefont
  {Burgd{\"o}rfer}, \citenamefont {Dunning}, \citenamefont {Sadeghpour},
  \citenamefont {Demler},\ and\ \citenamefont {Killian}}]{CamargoPRL2018}%
  \BibitemOpen
  \bibfield  {author} {\bibinfo {author} {\bibfnamefont {F.}~\bibnamefont
  {Camargo}}, \bibinfo {author} {\bibfnamefont {R.}~\bibnamefont {Schmidt}},
  \bibinfo {author} {\bibfnamefont {J.~D.}\ \bibnamefont {Whalen}}, \bibinfo
  {author} {\bibfnamefont {R.}~\bibnamefont {Ding}}, \bibinfo {author}
  {\bibfnamefont {J.}~\bibnamefont {Woehl}, \bibfnamefont {G.}}, \bibinfo
  {author} {\bibfnamefont {S.}~\bibnamefont {Yoshida}}, \bibinfo {author}
  {\bibfnamefont {J.}~\bibnamefont {Burgd{\"o}rfer}}, \bibinfo {author}
  {\bibfnamefont {F.~B.}\ \bibnamefont {Dunning}}, \bibinfo {author}
  {\bibfnamefont {H.~R.}\ \bibnamefont {Sadeghpour}}, \bibinfo {author}
  {\bibfnamefont {E.}~\bibnamefont {Demler}},\ and\ \bibinfo {author}
  {\bibfnamefont {T.~C.}\ \bibnamefont {Killian}},\ }\bibfield  {title}
  {\bibinfo {title} {Creation of rydberg polarons in a bose gas},\ }\href
  {https://doi.org/10.1103/PhysRevLett.120.083401} {\bibfield  {journal}
  {\bibinfo  {journal} {Phys. Rev. Lett.}\ }\textbf {\bibinfo {volume} {120}},\
  \bibinfo {pages} {083401} (\bibinfo {year} {2018})}\BibitemShut {NoStop}%
\bibitem [{\citenamefont {Astrakharchik}\ \emph {et~al.}(2021)\citenamefont
  {Astrakharchik}, \citenamefont {Pe{\~n}a~Ardila}, \citenamefont {Schmidt},
  \citenamefont {Jachymski},\ and\ \citenamefont
  {Negretti}}]{AstrakharchikCommPhys2021}%
  \BibitemOpen
  \bibfield  {author} {\bibinfo {author} {\bibfnamefont {G.~E.}\ \bibnamefont
  {Astrakharchik}}, \bibinfo {author} {\bibfnamefont {L.~A.}\ \bibnamefont
  {Pe{\~n}a~Ardila}}, \bibinfo {author} {\bibfnamefont {R.}~\bibnamefont
  {Schmidt}}, \bibinfo {author} {\bibfnamefont {K.}~\bibnamefont {Jachymski}},\
  and\ \bibinfo {author} {\bibfnamefont {A.}~\bibnamefont {Negretti}},\
  }\bibfield  {title} {\bibinfo {title} {Ionic polaron in a bose-einstein
  condensate},\ }\href {https://doi.org/10.1038/s42005-021-00597-1} {\bibfield
  {journal} {\bibinfo  {journal} {Communications Physics}\ }\textbf {\bibinfo
  {volume} {4}},\ \bibinfo {pages} {94} (\bibinfo {year} {2021})}\BibitemShut
  {NoStop}%
\bibitem [{\citenamefont {Pe{\~n}a~Ardila}\ and\ \citenamefont
  {Camacho-Guardian}(2025)}]{PenaArdilaCamacho2025}%
  \BibitemOpen
  \bibfield  {author} {\bibinfo {author} {\bibfnamefont {L.~A.}\ \bibnamefont
  {Pe{\~n}a~Ardila}}\ and\ \bibinfo {author} {\bibfnamefont {A.}~\bibnamefont
  {Camacho-Guardian}},\ }\bibfield  {title} {\bibinfo {title} {Polaronic
  dressing of bound states},\ }\href@noop {} {\bibfield  {journal} {\bibinfo
  {journal} {Phys. Rev. A}\ }\textbf {\bibinfo {volume} {111}},\ \bibinfo
  {pages} {L061302} (\bibinfo {year} {2025})}\BibitemShut {NoStop}%
\bibitem [{\citenamefont {Tempere}\ \emph {et~al.}(2009)\citenamefont
  {Tempere}, \citenamefont {Casteels}, \citenamefont {Oberthaler},
  \citenamefont {Knoop}, \citenamefont {Timmermans},\ and\ \citenamefont
  {Devreese}}]{TemperePRB2009}%
  \BibitemOpen
  \bibfield  {author} {\bibinfo {author} {\bibfnamefont {J.}~\bibnamefont
  {Tempere}}, \bibinfo {author} {\bibfnamefont {W.}~\bibnamefont {Casteels}},
  \bibinfo {author} {\bibfnamefont {M.~K.}\ \bibnamefont {Oberthaler}},
  \bibinfo {author} {\bibfnamefont {S.}~\bibnamefont {Knoop}}, \bibinfo
  {author} {\bibfnamefont {E.}~\bibnamefont {Timmermans}},\ and\ \bibinfo
  {author} {\bibfnamefont {J.~T.}\ \bibnamefont {Devreese}},\ }\bibfield
  {title} {\bibinfo {title} {Feynman path-integral treatment of the
  {BEC}-impurity polaron},\ }\href {https://doi.org/10.1103/PhysRevB.80.184504}
  {\bibfield  {journal} {\bibinfo  {journal} {Phys. Rev. B}\ }\textbf {\bibinfo
  {volume} {80}},\ \bibinfo {pages} {184504} (\bibinfo {year}
  {2009})}\BibitemShut {NoStop}%
\bibitem [{\citenamefont {Rath}\ and\ \citenamefont
  {Schmidt}(2013)}]{RathSchmidtPRA2013}%
  \BibitemOpen
  \bibfield  {author} {\bibinfo {author} {\bibfnamefont {S.~P.}\ \bibnamefont
  {Rath}}\ and\ \bibinfo {author} {\bibfnamefont {R.}~\bibnamefont {Schmidt}},\
  }\bibfield  {title} {\bibinfo {title} {Field-theoretical study of the {Bose}
  polaron},\ }\href {https://doi.org/10.1103/PhysRevA.88.053632} {\bibfield
  {journal} {\bibinfo  {journal} {Phys. Rev. A}\ }\textbf {\bibinfo {volume}
  {88}},\ \bibinfo {pages} {053632} (\bibinfo {year} {2013})}\BibitemShut
  {NoStop}%
\bibitem [{\citenamefont {Shashi}\ \emph {et~al.}(2014)\citenamefont {Shashi},
  \citenamefont {Grusdt}, \citenamefont {Abanin},\ and\ \citenamefont
  {Demler}}]{ShashiPRA2014}%
  \BibitemOpen
  \bibfield  {author} {\bibinfo {author} {\bibfnamefont {A.}~\bibnamefont
  {Shashi}}, \bibinfo {author} {\bibfnamefont {F.}~\bibnamefont {Grusdt}},
  \bibinfo {author} {\bibfnamefont {D.~A.}\ \bibnamefont {Abanin}},\ and\
  \bibinfo {author} {\bibfnamefont {E.}~\bibnamefont {Demler}},\ }\bibfield
  {title} {\bibinfo {title} {Radio-frequency spectroscopy of polarons in
  ultracold {Bose} gases},\ }\href {https://doi.org/10.1103/PhysRevA.89.053617}
  {\bibfield  {journal} {\bibinfo  {journal} {Phys. Rev. A}\ }\textbf {\bibinfo
  {volume} {89}},\ \bibinfo {pages} {053617} (\bibinfo {year}
  {2014})}\BibitemShut {NoStop}%
\bibitem [{\citenamefont {Grusdt}\ \emph {et~al.}(2015)\citenamefont {Grusdt},
  \citenamefont {Shchadilova}, \citenamefont {Rubtsov},\ and\ \citenamefont
  {Demler}}]{GrusdtSciRep2015}%
  \BibitemOpen
  \bibfield  {author} {\bibinfo {author} {\bibfnamefont {F.}~\bibnamefont
  {Grusdt}}, \bibinfo {author} {\bibfnamefont {Y.~E.}\ \bibnamefont
  {Shchadilova}}, \bibinfo {author} {\bibfnamefont {A.~N.}\ \bibnamefont
  {Rubtsov}},\ and\ \bibinfo {author} {\bibfnamefont {E.}~\bibnamefont
  {Demler}},\ }\bibfield  {title} {\bibinfo {title} {Renormalization group
  approach to the {Fr{\"o}hlich} polaron model: application to impurity-{BEC}
  problem},\ }\href {https://doi.org/10.1038/srep12124} {\bibfield  {journal}
  {\bibinfo  {journal} {Sci. Rep.}\ }\textbf {\bibinfo {volume} {5}},\ \bibinfo
  {pages} {12124} (\bibinfo {year} {2015})}\BibitemShut {NoStop}%
\bibitem [{\citenamefont {Jaksch}\ \emph {et~al.}(1998)\citenamefont {Jaksch},
  \citenamefont {Bruder}, \citenamefont {Cirac}, \citenamefont {Gardiner},\
  and\ \citenamefont {Zoller}}]{JakschPRL1998}%
  \BibitemOpen
  \bibfield  {author} {\bibinfo {author} {\bibfnamefont {D.}~\bibnamefont
  {Jaksch}}, \bibinfo {author} {\bibfnamefont {C.}~\bibnamefont {Bruder}},
  \bibinfo {author} {\bibfnamefont {J.~I.}\ \bibnamefont {Cirac}}, \bibinfo
  {author} {\bibfnamefont {C.~W.}\ \bibnamefont {Gardiner}},\ and\ \bibinfo
  {author} {\bibfnamefont {P.}~\bibnamefont {Zoller}},\ }\bibfield  {title}
  {\bibinfo {title} {Cold bosonic atoms in optical lattices},\ }\href
  {https://doi.org/10.1103/PhysRevLett.81.3108} {\bibfield  {journal} {\bibinfo
   {journal} {Phys. Rev. Lett.}\ }\textbf {\bibinfo {volume} {81}},\ \bibinfo
  {pages} {3108} (\bibinfo {year} {1998})}\BibitemShut {NoStop}%
\bibitem [{\citenamefont {Fisher}\ \emph {et~al.}(1989)\citenamefont {Fisher},
  \citenamefont {Weichman}, \citenamefont {Grinstein},\ and\ \citenamefont
  {Fisher}}]{PhysRevB.40.546}%
  \BibitemOpen
  \bibfield  {author} {\bibinfo {author} {\bibfnamefont {M.~P.~A.}\
  \bibnamefont {Fisher}}, \bibinfo {author} {\bibfnamefont {P.~B.}\
  \bibnamefont {Weichman}}, \bibinfo {author} {\bibfnamefont {G.}~\bibnamefont
  {Grinstein}},\ and\ \bibinfo {author} {\bibfnamefont {D.~S.}\ \bibnamefont
  {Fisher}},\ }\bibfield  {title} {\bibinfo {title} {Boson localization and the
  superfluid-insulator transition},\ }\href
  {https://doi.org/10.1103/PhysRevB.40.546} {\bibfield  {journal} {\bibinfo
  {journal} {Phys. Rev. B}\ }\textbf {\bibinfo {volume} {40}},\ \bibinfo
  {pages} {546} (\bibinfo {year} {1989})}\BibitemShut {NoStop}%
\bibitem [{\citenamefont {Capogrosso-Sansone}\ \emph
  {et~al.}(2007)\citenamefont {Capogrosso-Sansone}, \citenamefont {Prokof'ev},\
  and\ \citenamefont {Svistunov}}]{PhysRevB.75.134302}%
  \BibitemOpen
  \bibfield  {author} {\bibinfo {author} {\bibfnamefont {B.}~\bibnamefont
  {Capogrosso-Sansone}}, \bibinfo {author} {\bibfnamefont {N.~V.}\ \bibnamefont
  {Prokof'ev}},\ and\ \bibinfo {author} {\bibfnamefont {B.~V.}\ \bibnamefont
  {Svistunov}},\ }\bibfield  {title} {\bibinfo {title} {Phase diagram and
  thermodynamics of the three-dimensional bose-hubbard model},\ }\href
  {https://doi.org/10.1103/PhysRevB.75.134302} {\bibfield  {journal} {\bibinfo
  {journal} {Phys. Rev. B}\ }\textbf {\bibinfo {volume} {75}},\ \bibinfo
  {pages} {134302} (\bibinfo {year} {2007})}\BibitemShut {NoStop}%
\bibitem [{\citenamefont {Capogrosso-Sansone}\ \emph
  {et~al.}(2008)\citenamefont {Capogrosso-Sansone}, \citenamefont {S{\"o}yler},
  \citenamefont {Prokof'ev},\ and\ \citenamefont
  {Svistunov}}]{CapogrossoSansonePRA2008}%
  \BibitemOpen
  \bibfield  {author} {\bibinfo {author} {\bibfnamefont {B.}~\bibnamefont
  {Capogrosso-Sansone}}, \bibinfo {author} {\bibfnamefont {{\c S}.~G.}\
  \bibnamefont {S{\"o}yler}}, \bibinfo {author} {\bibfnamefont {N.~V.}\
  \bibnamefont {Prokof'ev}},\ and\ \bibinfo {author} {\bibfnamefont {B.~V.}\
  \bibnamefont {Svistunov}},\ }\bibfield  {title} {\bibinfo {title} {Phase
  diagram and thermodynamics of the two-dimensional bose-hubbard model},\
  }\href {https://doi.org/10.1103/PhysRevA.77.015602} {\bibfield  {journal}
  {\bibinfo  {journal} {Phys. Rev. A}\ }\textbf {\bibinfo {volume} {77}},\
  \bibinfo {pages} {015602} (\bibinfo {year} {2008})}\BibitemShut {NoStop}%
\bibitem [{\citenamefont {Kiely}\ \emph {et~al.}(2025)\citenamefont {Kiely},
  \citenamefont {Zhang},\ and\ \citenamefont
  {Mueller}}]{KielyZhangMuellerPRA2025}%
  \BibitemOpen
  \bibfield  {author} {\bibinfo {author} {\bibfnamefont {T.~G.}\ \bibnamefont
  {Kiely}}, \bibinfo {author} {\bibfnamefont {C.}~\bibnamefont {Zhang}},\ and\
  \bibinfo {author} {\bibfnamefont {E.~J.}\ \bibnamefont {Mueller}},\
  }\bibfield  {title} {\bibinfo {title} {Vacancy-assisted superfluid drag},\
  }\href {https://doi.org/10.1103/PhysRevA.111.053302} {\bibfield  {journal}
  {\bibinfo  {journal} {Phys. Rev. A}\ }\textbf {\bibinfo {volume} {111}},\
  \bibinfo {pages} {053302} (\bibinfo {year} {2025})}\BibitemShut {NoStop}%
\bibitem [{\citenamefont {Zhang}(2026{\natexlab{a}})}]{chaoletter}%
  \BibitemOpen
  \bibfield  {author} {\bibinfo {author} {\bibfnamefont {C.}~\bibnamefont
  {Zhang}},\ }\bibfield  {title} {\bibinfo {title} {Impurity self-trapping in
  lattice bose systems},\ }\href {https://arxiv.org/abs/2601.11058} {\bibfield
  {journal} {\bibinfo  {journal} {arXiv:2601.11058}\ } (\bibinfo {year}
  {2026}{\natexlab{a}})}\BibitemShut {NoStop}%
\bibitem [{\citenamefont {Zhang}(2026{\natexlab{b}})}]{chaoPRB}%
  \BibitemOpen
  \bibfield  {author} {\bibinfo {author} {\bibfnamefont {C.}~\bibnamefont
  {Zhang}},\ }\bibfield  {title} {\bibinfo {title} {Mobile impurity coupled to
  correlated lattice bosons},\ }\href {https://arxiv.org/abs/2601.11062}
  {\bibfield  {journal} {\bibinfo  {journal} {arXiv: 2601.11062}\ } (\bibinfo
  {year} {2026}{\natexlab{b}})}\BibitemShut {NoStop}%
\bibitem [{\citenamefont {Santiago-García}\ \emph {et~al.}(2024)\citenamefont
  {Santiago-García}, \citenamefont {Castillo-López},\ and\ \citenamefont
  {Camacho-Guardian}}]{Santiago-García_2024}%
  \BibitemOpen
  \bibfield  {author} {\bibinfo {author} {\bibfnamefont {M.}~\bibnamefont
  {Santiago-García}}, \bibinfo {author} {\bibfnamefont {S.~G.}\ \bibnamefont
  {Castillo-López}},\ and\ \bibinfo {author} {\bibfnamefont {A.}~\bibnamefont
  {Camacho-Guardian}},\ }\bibfield  {title} {\bibinfo {title} {Lattice polaron
  in a bose–einstein condensate of hard-core bosons},\ }\href
  {https://doi.org/10.1088/1367-2630/ad503e} {\bibfield  {journal} {\bibinfo
  {journal} {New Journal of Physics}\ }\textbf {\bibinfo {volume} {26}},\
  \bibinfo {pages} {063015} (\bibinfo {year} {2024})}\BibitemShut {NoStop}%
\bibitem [{\citenamefont {Prokof’ev}\ \emph {et~al.}(1998)\citenamefont
  {Prokof’ev}, \citenamefont {Svistunov},\ and\ \citenamefont
  {Tupitsyn}}]{ProkofevJETP1998}%
  \BibitemOpen
  \bibfield  {author} {\bibinfo {author} {\bibfnamefont {N.~V.}\ \bibnamefont
  {Prokof’ev}}, \bibinfo {author} {\bibfnamefont {B.~V.}\ \bibnamefont
  {Svistunov}},\ and\ \bibinfo {author} {\bibfnamefont {I.~S.}\ \bibnamefont
  {Tupitsyn}},\ }\bibfield  {title} {\bibinfo {title} {Worm algorithm in
  quantum monte carlo simulations},\ }\href@noop {} {\bibfield  {journal}
  {\bibinfo  {journal} {JETP}\ }\textbf {\bibinfo {volume} {87}},\ \bibinfo
  {pages} {310} (\bibinfo {year} {1998})}\BibitemShut {NoStop}%
\bibitem [{\citenamefont {Prokof'ev}\ \emph {et~al.}(1998)\citenamefont
  {Prokof'ev}, \citenamefont {Svistunov},\ and\ \citenamefont
  {Tupitsyn}}]{secondworm}%
  \BibitemOpen
  \bibfield  {author} {\bibinfo {author} {\bibfnamefont {N.~V.}\ \bibnamefont
  {Prokof'ev}}, \bibinfo {author} {\bibfnamefont {B.~V.}\ \bibnamefont
  {Svistunov}},\ and\ \bibinfo {author} {\bibfnamefont {I.~S.}\ \bibnamefont
  {Tupitsyn}},\ }\bibfield  {title} {\bibinfo {title} {“worm” algorithm in
  quantum monte carlo simulations},\ }\href
  {https://doi.org/https://doi.org/10.1016/S0375-9601(97)00957-2} {\bibfield
  {journal} {\bibinfo  {journal} {Physics Letters A}\ }\textbf {\bibinfo
  {volume} {238}},\ \bibinfo {pages} {253} (\bibinfo {year}
  {1998})}\BibitemShut {NoStop}%
\bibitem [{\citenamefont {Capogrosso-Sansone}\ \emph
  {et~al.}(2010)\citenamefont {Capogrosso-Sansone}, \citenamefont {S\"oyler},
  \citenamefont {Prokof'ev},\ and\ \citenamefont
  {Svistunov}}]{PhysRevA.81.053622}%
  \BibitemOpen
  \bibfield  {author} {\bibinfo {author} {\bibfnamefont {B.}~\bibnamefont
  {Capogrosso-Sansone}}, \bibinfo {author} {\bibfnamefont {G.}~\bibnamefont
  {S\"oyler}}, \bibinfo {author} {\bibfnamefont {N.~V.}\ \bibnamefont
  {Prokof'ev}},\ and\ \bibinfo {author} {\bibfnamefont {B.~V.}\ \bibnamefont
  {Svistunov}},\ }\bibfield  {title} {\bibinfo {title} {Critical entropies for
  magnetic ordering in bosonic mixtures on a lattice},\ }\href
  {https://doi.org/10.1103/PhysRevA.81.053622} {\bibfield  {journal} {\bibinfo
  {journal} {Phys. Rev. A}\ }\textbf {\bibinfo {volume} {81}},\ \bibinfo
  {pages} {053622} (\bibinfo {year} {2010})}\BibitemShut {NoStop}%
\bibitem [{\citenamefont {Lingua}\ \emph {et~al.}(2018)\citenamefont {Lingua},
  \citenamefont {Capogrosso-Sansone}, \citenamefont {Safavi-Naini},
  \citenamefont {Jahangiri},\ and\ \citenamefont {Penna}}]{Lingua_2018}%
  \BibitemOpen
  \bibfield  {author} {\bibinfo {author} {\bibfnamefont {F.}~\bibnamefont
  {Lingua}}, \bibinfo {author} {\bibfnamefont {B.}~\bibnamefont
  {Capogrosso-Sansone}}, \bibinfo {author} {\bibfnamefont {A.}~\bibnamefont
  {Safavi-Naini}}, \bibinfo {author} {\bibfnamefont {A.~J.}\ \bibnamefont
  {Jahangiri}},\ and\ \bibinfo {author} {\bibfnamefont {V.}~\bibnamefont
  {Penna}},\ }\bibfield  {title} {\bibinfo {title} {Multiworm algorithm quantum
  monte carlo},\ }\href {https://doi.org/10.1088/1402-4896/aadd7a} {\bibfield
  {journal} {\bibinfo  {journal} {Physica Scripta}\ }\textbf {\bibinfo {volume}
  {93}},\ \bibinfo {pages} {105402} (\bibinfo {year} {2018})}\BibitemShut
  {NoStop}%
\bibitem [{\citenamefont {Fukuhara}\ \emph {et~al.}(2013)\citenamefont
  {Fukuhara}, \citenamefont {Schau\ss{}}, \citenamefont {Endres}, \citenamefont
  {Hild}, \citenamefont {Cheneau}, \citenamefont {Bloch},\ and\ \citenamefont
  {Gross}}]{FukuharaNatPhys2013}%
  \BibitemOpen
  \bibfield  {author} {\bibinfo {author} {\bibfnamefont {T.}~\bibnamefont
  {Fukuhara}}, \bibinfo {author} {\bibfnamefont {P.}~\bibnamefont
  {Schau\ss{}}}, \bibinfo {author} {\bibfnamefont {M.}~\bibnamefont {Endres}},
  \bibinfo {author} {\bibfnamefont {S.}~\bibnamefont {Hild}}, \bibinfo {author}
  {\bibfnamefont {M.}~\bibnamefont {Cheneau}}, \bibinfo {author} {\bibfnamefont
  {I.}~\bibnamefont {Bloch}},\ and\ \bibinfo {author} {\bibfnamefont
  {C.}~\bibnamefont {Gross}},\ }\bibfield  {title} {\bibinfo {title} {Quantum
  dynamics of a mobile spin impurity},\ }\href
  {https://doi.org/10.1038/nphys2561} {\bibfield  {journal} {\bibinfo
  {journal} {Nat. Phys.}\ }\textbf {\bibinfo {volume} {9}},\ \bibinfo {pages}
  {235} (\bibinfo {year} {2013})}\BibitemShut {NoStop}%
\end{thebibliography}%
\end{document}